\documentclass[fleqn,usenatbib]{mnras}
\newcommand\bm{\boldsymbol}
\usepackage{newtxtext,newtxmath}

\usepackage[T1]{fontenc}

\DeclareRobustCommand{\VAN}[3]{#2}
\let\VANthebibliography\thebibliography
\def\thebibliography{\DeclareRobustCommand{\VAN}[3]{##3}\VANthebibliography}

\usepackage{graphicx}	
\usepackage{amsmath}	

\title[FLOWER]{Data-driven dust inference at mid-to-high Galactic latitudes using probabilistic machine learning}

\author[M. O'Callaghan et al.]{
Matthew O'Callaghan,$^{1}$\thanks{E-mail:mo503@cam.ac.uk}
Kaisey S. Mandel,$^{1,2,3}$
Gerry Gilmore$^{1,4}$
\\
$^{1}$Institute of Astronomy, University of Cambridge, Madingley Road, Cambridge, CB3 0HA, UK\\
$^{2}$Statistical Laboratory, DPMMS, University of Cambridge, Wilberforce Road, Cambridge, CB3 0WB, UK\\
$^{3}$Kavli Institute for Cosmology, University of Cambridge, Madingley Road, Cambridge, CB3 0HA, UK\\
$^{4}$Institute of Astrophysics, FORTH, Crete, Greece\\}

\date{Accepted XXX. Received YYY; in original form ZZZ}

\pubyear{2025}

\begin{document}
\label{firstpage}
\pagerange{\pageref{firstpage}--\pageref{lastpage}}
\maketitle

\begin{abstract}
We present a method for accurately and precisely inferring photometric dust reddening towards stars at mid-to-high Galactic latitudes ($|b| > 20^\circ$), using probabilistic machine learning to model the colour–magnitude distribution of zero-extinction stars in these regions. Photometric dust maps rely on a robust method for inferring stellar reddening. At high Galactic latitudes, where extinction is low, such inferences are particularly susceptible to contamination from modelling errors and prior assumptions, potentially introducing artificial structure into dust maps. In this work, we demonstrate the use of normalising flows to learn the conditional probability distribution of the photometric colour–magnitude relations of zero-extinction stars, conditioned on Galactic cylindrical coordinates for stars 
at mid-to-high Galactic latitudes. By using the normalising flow to model the colour–magnitude diagram, we infer the posterior distribution of dust extinction towards stars along different lines of sight by marginalising over the {colour-magnitude} flow. We validate our method using data from Gaia, Pan-STARRS, and 2MASS, showing that we can recover unbiased posteriors and successfully detect dust along the line-of-sight in two calibration regions at mid-Galactic latitude that have been extensively studied in the context of polarisation surveys.

\end{abstract}




\section{Introduction}
\begin{figure*}
    \centering
    \includegraphics[width=\textwidth]{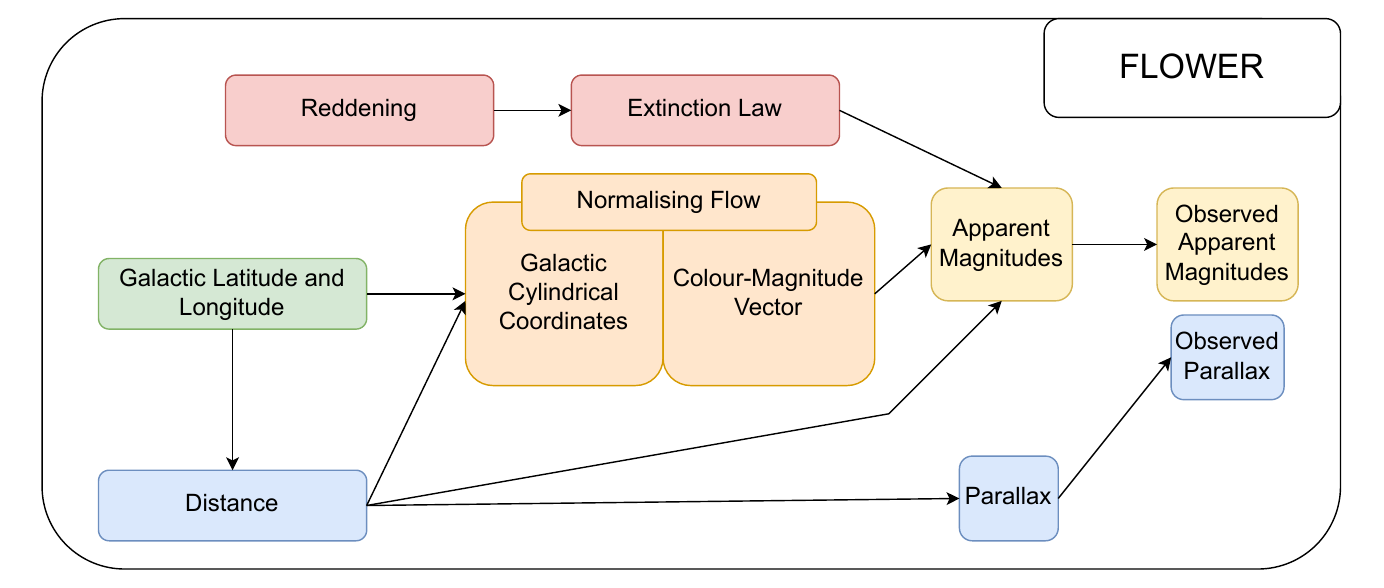}
        \caption{FLOW-based Estimates of Reddening, a method which learns the zero-extinction colour-magnitude relations as a probability distribution conditional on the Galactic cylindrical coordinates. This diagram provides an overview of the forward model we developed to describe photometric colour-magnitude vectors at high Galactic latitudes. Omitted from this diagram is the forward modelling of selection effects.}
    \label{fig:flower}
\end{figure*}

We present FLOW-based Estimation of Reddening (FLOWER; Figure \ref{fig:flower}), a novel method for inferring photometric dust reddening by learning the colour-magnitude probability distribution of high Galactic latitude zero-extinction stars using normalising flows. After the normalising flow has been trained, we infer the photometric {dust} reddening of stars at Galactic latitudes $|b|>20^\circ$. The technical details of our model are described in Section \ref{methods}. The description of the surveys used throughout the paper can be found in Section \ref{data_section}, and details of the zero-extinction datasets used to train our model can be found in Section \ref{training_regions}. After training our model, we infer line-of-sight {(LOS)} reddening for different calibration regions and outline the results in Section \ref{inference_results}.
\subsection{The diffuse interstellar medium at mid-to-high Galactic latitudes}
Charting the small-scale variations of the dusty interstellar medium (ISM) in the Milky Way has consequences across a broad range of applications in astronomy and cosmology. At mid- to high-Galactic latitudes, where the ISM is more diffuse than towards the Galactic centre, the spatial structure of the ISM is dominated by small-scale diffuse structures such as filaments (\citealt{10.1093/mnras/stae2274}; \citealt{kalberla2025cold}). The diffuse structure of the ISM at high latitudes remains a foreground contaminant for many astronomical and cosmological applications (\citealt{2024MNRAS.527.5751A}; \citealt{2025JCAP...02..045K}), and to fully leverage the data from future surveys for scientific applications, it is necessary to account for the ISM in a reliable way that minimises systematic effects and spurious correlations with large-scale structure.

Astronomers rely on dust maps to account for the effects of the ISM on observations. Through absorption, re-emission, and scattering, dust grains {cause the attenuation} of starlight preferentially at shorter wavelengths, {inducing} a reddening effect for the observer \citep{2001ApJ...554..778L}. The observable properties of dust have been used to characterise and map the ISM. {Emission} has long been established as a method for inferring the LOS integrated 2D distribution of dust, either by measuring the dust column density using far infrared (IR) emission (\citealt{SFD}; \citealt{planck_dust_cmb}) or neutral hydrogen emission \citep{Lenz_2017} {under the assumption of well-mixing of gas and dust}. Although these methods have provided dust maps on fine angular scales, they are restricted to 2D reconstructions of the ISM. {Early attempts to characterise the three-dimensional structure of the dusty ISM were carried out using photometry \citep{2006A&A...453..635M}.} The advent of a large dataset of parallax measurements from the European Space Agency's Gaia mission \citep{GAIA_MISSION} has allowed astronomers to leverage photometric reddening to infer the 3D structure of the dusty ISM in our Galaxy {to a much higher resolution} (\citealt{greenmap}; \citealt{mlgreen}; \citealt{edenhofer23}; \citealt{gspphot}).

{An effort} must be made to construct 3D dust maps in a bias-free manner to ensure their usefulness in accounting for the effects of the ISM on observation. However, most emission and reddening {or extinction-based} dust maps exhibit significant correlation with the large-scale structure of the Galaxy \citep{2019ApJ...870..120C} except the \citealt{Lenz_2017} HI emission dust map. In integrated emission-based maps, this correlation often arises from {imperfect} removal of
the cosmic infrared background (CIB) and the cumulative emission of dusty star-forming galaxies \citep{Mudur_2023}. For {reddening-based} maps, the imperfect removal of galaxies and QSOs from stellar
catalogues \citep{Mudur_2023}, strong priors in prior-dominant regions, and modelling inaccuracies \citep{high_gal_me} can lead to a systematic bias of the extinction posterior correlated with large-scale {and Galactic} structure. Integrated, reddening-based dust maps which do not invoke the large-scale structure and do not contain extra-Galactic objects are promising in this context.

The significant spatial structure of the diffuse ISM is particularly important in the context of CMB experiments \citep{Planck}. A better understanding of the 2D \citep{2024MNRAS.527.5751A} and 3D \citep{2022arXiv221107667Z} distribution of the dusty ISM in a bias-free manner is necessary for the next generation of CMB experiments, such as the Simons Observatory \citep{2021AAS...23721403X} and CMB-S4 \citep{2022ApJ...926...54A}, particularly in the pursuit of primordial B-modes \citep{Planck}. Although dust emits at CMB frequencies (around $150$ GHz), its emission dominates at higher frequencies (around $350$ GHz). To account for the dusty ISM foreground in primordial B-mode experiments, astronomers map the ISM at the dominant dust emission frequencies and extrapolate {it} down to CMB frequencies \citep{planck_dust_cmb} using an empirical dust emission model.

Overly simplistic assumptions {on} the spatial variation of dust can cause decorrelation {when extrapolating} across frequencies {\citep{2023MNRAS.519.4370M}. In particular, if the dust has different compositions, temperatures, and if the polarisation angle is misaligned for clouds along the LOS \citep{2015MNRAS.451L..90T}}. Although multiple LOS components may exist in thermal emission, single-component approximations are still widely used \citep{2025ApJS..276...45L}. A probabilistic description of the 3D structure of the dusty ISM is necessary in this context \citep{2023MNRAS.519.4370M}. At high Galactic latitudes, the Intermediate Velocity Clouds (IVCs) are the most distant clouds with detectable dust content, which will influence the next generation of CMB experiments \citep{tassis2018pasiphae}, providing a lower bound of $2$ kpc on the desired reach of reddening-based dust maps for the LOS decorrelation of the ISM. 

To properly constrain extinction at high Galactic latitudes in the context of fine-scale variation, methods which deal with photometric extinction and dust in a probabilistic way that is not overconfident are essential. In this paper, we present a flexible, data-driven, and stellar parameter-independent method for determining the reddening of stars at mid- to high-Galactic latitudes.

\subsection{Current methods for inferring photometric dust reddening}
{Reddening-based} dust maps have delivered success in generating large-scale, 3D {dust} maps in our Galaxy (\citealt{greenmap}; \citealt{mlgreen}; \citealt{edenhofer23}; \citealt{gspphot}). All of these maps {rely on} a precise and accurate method for inferring the photometric {reddening towards} an individual star. The main subject of this paper is the \textbf{reddening estimate} for a given star, {which is} the first building block of dust maps. We describe a method suited to inferring photometric {reddening} at mid- to high-Galactic latitudes by using {probabilistic machine learning} to learn the probability distribution of the photometric colours of zero-extinction stars in these regions. We discuss the current methods for inferring reddening {(or extinction)}, and then we discuss the proposed changes that we have implemented.

Reddening is inferred by assuming some zero-extinction sample of stars in photometric or spectroscopic space using either ab initio stellar models or a subsample of observed stars (\citealt{Green_2014}; \citealt{Lallement}; \citealt{gspphot}). After assuming (or jointly inferring{;} \citealt{mlgreen}) an extinction law and adopting astrophysical priors, the observed stars are compared against the underlying zero-extinction sample and the {dust extinction} is inferred. Defining the zero-extinction sample using stellar evolution models presents many difficulties, including unknown model inaccuracies, the impact of the prior on extinction, and parameter degeneracies, which pose problems particularly at high Galactic latitudes, where dust is diffuse \citep{high_gal_me}. These ab initio methods (such as {\citealt{starhorse,gspphot}}) are useful as they allow us to test our understanding of the underlying astrophysical models and incorporate known astrophysics into the inference. 

\begin{figure}
    \centering
    \includegraphics[width=0.5\textwidth]{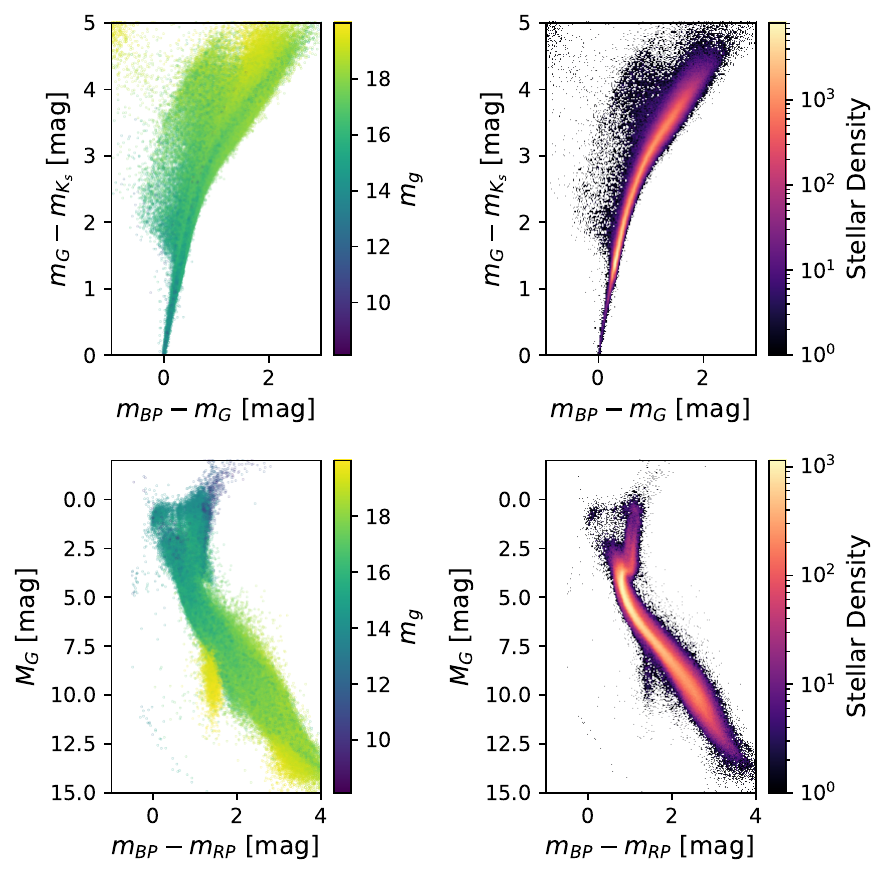}
        \caption{Gaia-2MASS colour-colour and colour-magnitude relations for zero-extinction stars with the Gaia G apparent magnitude (left column) and the stellar density (right column) as the hue. We model the probability distribution of these relations, which is to be marginalised over to obtain extinction posteriors. A simple functional fit of the colour-colour and colour-magnitude relations cannot capture the genuine astrophysical scatter that occurs observationally. Moreover, there are astrophysical systematics in the stellar parameters of stars that change as a function of Galactic cylindrical coordinates. Without modelling the scatter and the systematic change in stellar population, the extinction posterior will be biased at different points of the Galaxy.}
    \label{fig:colour_colour}
\end{figure}

Using a data-driven approach to define the zero-extinction sample relies on choosing a subset of stars that lie behind a negligible amount of dust (see Figure \ref{fig:colour_colour} for an example sample), which is assumed to be devoid of extinction. Current photometric dust maps and their downstream data-driven methods trained on them {\citep{Green_2015,Green_2014,Green_2018,greenmap,Lallement,mlgreen,edenhofer23}} derive (or are trained on) {extinction or reddening values that are} calculated by first fitting zero-extinction photometry as a \textit{deterministic function} of other stellar parameters or colours. Prior distributions are then adopted over these stellar parameters, and the photometric dust reddening is inferred for another test set of stars using inference methods.

Traditional methods usually learn the colour-colour relations as a deterministic function of stellar parameters. These are useful for large-scale dust maps as the underlying population of stars changes significantly from the training set of stars at high Galactic latitudes. However, issues arise from modelling the colour-colour relations as a deterministic function that can be solved specifically at high Galactic latitudes. When using the usual deterministic method for inferring the colour-colour relations, priors over the stellar parameters and spatial correlations of dust are adopted. However, at high Galactic latitudes, extinction is highly prior dependent, and choosing incorrect priors can bias the extinction estimates \citep{high_gal_me}. Moreover, the colour-colour relations are not a perfect one-dimensional curve and have both observational and astrophysical scatter as a function of the {astrophysical parameters}. Furthermore, it is difficult to map stellar type perfectly onto these colour-colour relations due to degeneracies in stellar parameters, particularly at very bright and very faint absolute magnitudes. 

Some of these modelling issues can be seen in a sample of low-extinction stars presented in Figure \ref{fig:colour_colour}. A functional fit of the colour-colour relations can miss the astrophysical complexities of the data. Moreover, cleaning this sample or fitting a functional relation can introduce bias and cause stars with genuinely zero extinction to appear as {extinguished} in the inference.
\subsection{Leveraging probabilistic machine learning to learn zero-extinction stars assuming an exponential disk model of the Galaxy}

We propose an alternative method for including the zero-extinction sample of stars to infer the reddening of stars at mid- and high-Galactic latitude regions that can overcome some of the issues in the traditional methods. Our method is based on a single assumption: that the \textit{intrinsic} colour-magnitude relations of stars at high Galactic latitudes are conditional mainly on the Galactic cylindrical coordinates. 

This assumption is founded on a three-component exponential disk model for the Milky Way, which models the stellar populations conditional on the absolute perpendicular distance above the Galactic plane, $Z\equiv|Z|$, and the radial distance from the Galactic centre, $R$. Evidence of the Galactic disk structure dates back to the discovery of the thin and thick disk (\citealt{1982PASJ...34..365Y}; \citealt{1983MNRAS.202.1025G}; \citealt{2008ApJ...673..864J}; \citealt{2025arXiv250717629I}), where the two disks were found to differ in their density, kinematic properties, age distributions and chemical abundances. In the solar neighbourhood, these changes are mainly correlated with Galactic height $Z$, where the density profile is exponential with a scale height that exhibits a Galactic radial dependence \citep{2025arXiv250717629I}. The population change will have an impact on the typical photometry one expects to see at different distances, and to reliably learn the colour-magnitude relations, it is essential that we properly account for the stellar population change to not introduce a systematic {that will bias} our LOS extinction profile.

Stellar populations are complex and often cannot be characterised in an unbiased manner from simple selection cuts on the photometry. In Figure \ref{fig:training_region_hr} we display an example colour-magnitude diagram for a {region} of high Galactic latitude stars. It can be difficult to introduce cuts to the sample as they may bias the selection function and, subsequently, the extinction posterior in a non-trivial manner. Other than quality cuts, we make very few ad-hoc adjustments to the underlying data so as not to bias the data in a manner that is difficult to account for. Without any informed, unbiased cuts, we have no reason a priori to exclude many of these sources.

\begin{figure}
    \centering
    \includegraphics[width=0.5\textwidth]{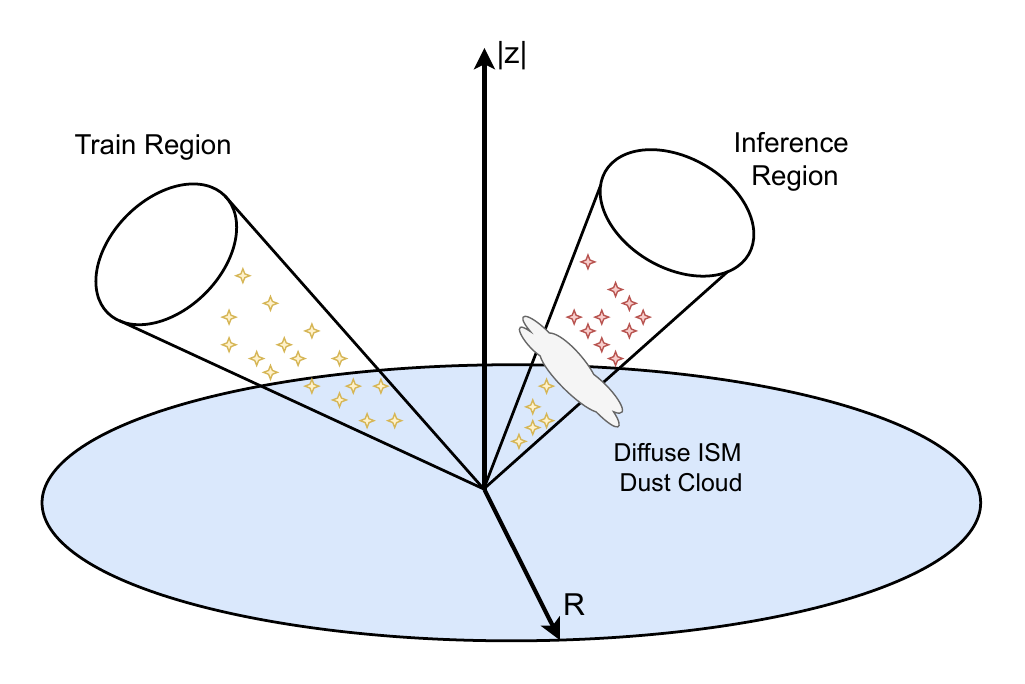}
        \caption{Sketch diagram showing the basic assumption of our model: that nearby lines-of-sight at mid-to high-Galactic latitude should have a similar probability distribution of colours conditional on the Galactic cylindrical coordinates.}
    \label{fig:training_inference}
\end{figure}

\begin{figure}
    \centering
    \includegraphics[width=0.5\textwidth]{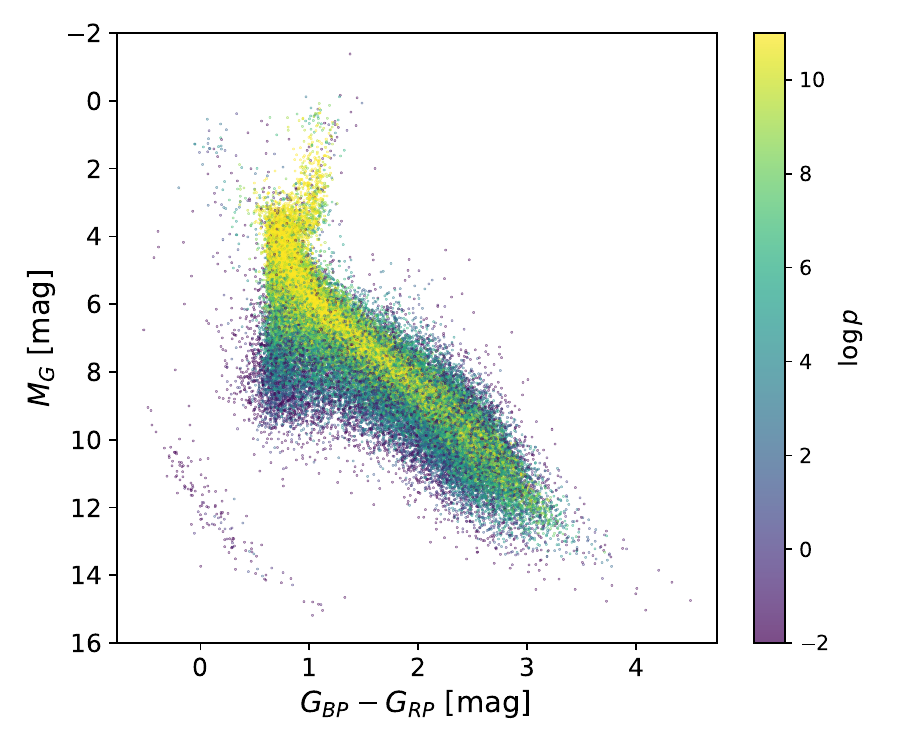}
        \caption{Gaia $G$ band absolute magnitude using the \citealt{Bailer_Jones_distance} distance estimate versus $BP-RP$ colour for the selection of high Galactic latitude stars used as the training sample in this paper. In our training sample, we make very few non-quality cuts to the data, as their impact on the selection function is non-trivial and they can bias the underlying sample significantly, which will affect the final reddening posterior. The hue indicates the log-probability density of the trained normalising flow on the data, where we can see that outlying data appear as low probability density. }
    \label{fig:training_region_hr}
\end{figure}

We learn the probability distribution $p_{\boldsymbol{\theta}}(\boldsymbol{x}|\, R, Z)$ of the colour-magnitude vector $\boldsymbol{x}$ parametrized by $\boldsymbol{\theta}$, conditional on the Galactic cylindrical coordinates using a normalising flow \citep{Kobyzev_2021}. Thus, we replace functionally learning the zero-extinction sample with learning the probability distribution of zero-extinction stars at mid- to high-Galactic latitudes. Assuming that we have enough zero-extinction stars in our training sample, we can infer the reddening of stars using Bayesian inference in a different LOS provided we have enough coverage of $Z$ and $R$ in our training set (see Figure \ref{fig:training_inference}).

We summarise the main benefits of our model as follows. \begin{enumerate}
    \item We learn the probability distribution of the colour-magnitude relations {given} Galactic cylindrical coordinates. This mitigates against modelling errors in matching photometry to spectroscopic labels and reliably defining prior distributions over those labels. Moreover, we implicitly integrate over the degeneracies with the other parameters when inferring reddening.
    \item Learning the colour-magnitude relations probabilistically as opposed to functionally allows us to account for their observational scatter and the genuine astrophysical deviation (Figure \ref{fig:colour_colour}). In particular, our proposed model makes no single-star assumption and, therefore, can account for binary stars.
    \item We implicitly learn the prior density of stars. Colloquially, we are learning {all catalogue knowledge observed about} the colour-magnitude relations given $Z$ and $R$ {in} our training sample.
\end{enumerate}

While we find that this allows us to account for many undesirable effects, our method is dependent on the selection function. {Careful consideration must be taken so that the selection function is consistent across different LOS under our model assumptions.}

\begin{figure}
    \centering
    \includegraphics[width=0.5\textwidth]{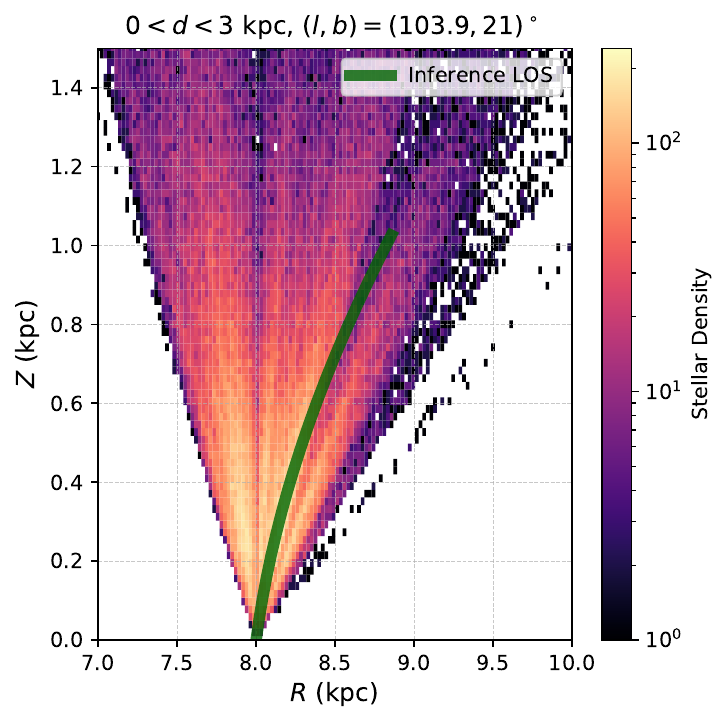}
        \caption{Coverage of Galactic Z and R in the training set used for training the normalising flow. This parameter space indicates the allowable $Z-R$ region for which one can reliably infer the reddening posterior. An inference line-of-sight should trace out a curve in this space that has good coverage from the training set if one wishes to satisfy the assumptions of the model. The green overlayed path corresponds to the one- and two-cloud inference regions lines-of-sight for $0<d<3$ kpc in the $Z-R$ space. This shows us that the coverage for this region is reliable and that we have significant star coverage from the training sample for these regions.}
    \label{fig:inference_pasiphae}
\end{figure}

\subsection{Related work}
This is not the first time normalising flows have been used to learn the colour-magnitude diagram. \citealt{cranmer2019modeling} have used normalising flows to model the Gaia colour-magnitude diagram, given estimates of the extinction from \citealt{bayestar} and data-driven stellar parameters. In this work, we use a sample of zero-extinction stars to learn intrinsic photometry as opposed to directly using estimates of extinction from other photometric surveys. {Other methods for learning photometry or spectra independent of stellar labels have been applied, for example \citealt{laroche2023closing} proposed an unsupervised generative
model for BP/RP spectra.}

\section{Methods}\label{methods}

We present a method to infer the reddening of stars at mid- to high-Galactic latitudes in a fast, precise, and accurate way. We do this by learning the conditional probability distribution of the colour-magnitude vector $\boldsymbol{x}$ of intrinsic zero-extinction stars, conditional on the Galactic absolute height $Z$, and radial coordinate $R$.
\subsection{Overview of model}
\subsubsection{Notation}
The distance modulus, the absolute Galactic height coordinate, and the Galactic radial coordinate are given by $\mu={5\log _{10}(d/10\text{pc})}$, \begin{equation}
    Z=d|\sin(\hat b)|{+Z_\odot},
\end{equation}
and \begin{equation}R=\sqrt{R_\odot^2+d^2\cos^2(\hat b)-2R_\odot d\cos(\hat b)\cos(\hat l)},\end{equation} respectively, where $d$ is the radial distance to the star in parsecs, $\hat b$ is the Galactic latitude in degrees, and $\hat l$ is the Galactic longitude, which we assume are observed exactly. 
We let $\boldsymbol{M}^{\rm int}$ denote the intrinsic absolute magnitudes. {We adopt $Z_\odot=25 \, \rm pc$ and $R_\odot=8 \, \rm kpc$}. Similar to \citealt{Green_2021}, we model the catalogue colour-magnitude vector as 
\begin{equation}\label{lambda_ref}
\begin{aligned}
    \bm{x} & = \boldsymbol{\Lambda}\boldsymbol{M}^{\rm int}
\end{aligned}
\end{equation}
where the absolute magnitudes $\boldsymbol{M}^{\rm int}$ are transformed by the square matrix \begin{equation}
    \boldsymbol{\Lambda} =
\begin{bmatrix}
1 & 0 & \cdots & 0 \\
-1 & 1 & \cdots & 0 \\
\vdots & \vdots & \ddots & \vdots \\
-1 & 0 &  \cdots & 1
\end{bmatrix}.
\end{equation}
For example, if $\boldsymbol{M}^{\rm int}=({M}_G^{\rm int},{M}_{K_s}^{\rm int})$ then 
$\bm{x}=({M}_G,{M}_{K_s}^{\rm int}-{M}_{G}^{\rm int})$. In the description of our model, we refer to \textit{the} absolute magnitude as $M=(\boldsymbol{M}^{\rm int})_1$.
\subsubsection{Model summary}
We define a Bayesian probabilistic model that is described in detail throughout this section, where the model consists of 
\begin{enumerate}
    \item $p(\bm{\hat m}|\, \boldsymbol{{x}},E,d)$, the likelihood of the intrinsic colour-magnitude vector $\boldsymbol{{x}}$, the reddening $E$, and distance $d$ given the observed apparent magnitudes $\bm{\hat m}$.
    \item $p{(\hat \omega} |\, d)$ the likelihood of the distance given the observed parallax $\hat \omega$.
    \item $p_{\boldsymbol{\theta}}(\boldsymbol{x} |\,R, Z)\approx p^*(\boldsymbol{x} |\,R, Z)$, parametrized by $\boldsymbol{\theta}
$, where the asterisks $p^*$ {indicates} the true, unknown distribution.
    \item $I(M|\,\mu)$ a density describing the Malmquist bias cut for the forward model given the faint and bright limits of the survey in the leading passband.
    \item $P(S=1| \, \boldsymbol{m}, \hat l , \hat b)$ the selection probability of the LOS binary selection variable $S$ given apparent magnitudes $\boldsymbol{m}$, for a given observed Galactic longitude $\hat l$ and latitude $\hat b$. {This is decoupled from the Malmquist bias, and} $M, d$ are needed {from} the forward model to calculate and evaluate the selection function on the apparent magnitude. 
    \item $p(d|\hat l, \hat b)$ a prior distribution on distance.
    \item $p(E)$ a prior distribution on reddening.
\end{enumerate}

\subsection{Likelihood functions for photometry and parallax}

The adopted likelihood function for the apparent magnitude components of the data vector is{:}
\begin{equation}
p(\bm{\hat m}|\,\boldsymbol{x}, E, d)=N(\bm{\hat m}|\, \boldsymbol{\Lambda}^{-1}\boldsymbol{x}+\boldsymbol{{R}}(E,C)E + \boldsymbol{1}\mu(d), \boldsymbol{\Sigma}),\end{equation}
where $\bm{\Sigma}$ is the error covariance matrix and $C$ is a photometric colour. The likelihood function for the parallax is given by
\begin{equation}P(\bm{\hat \omega} |\, d)=N\Big(\hat \omega{\, \rm{mas}} |\,\frac{1000}{d{\, \rm pc}} - \omega_{zp}{\, \rm{mas}},\sigma_\omega^2{\,\rm{mas}^2}\Big),\end{equation} 
where $\omega_{zp}$ is the zero-point offset for the parallax and $\sigma_\omega^2$ is the Gaussian measurement error on the parallax.

\subsection{The catalogue colour-magnitude probability distribution}
Strictly speaking, we are learning the \textit{catalogue distribution} of the colour-magnitude relations of stars at high Galactic latitudes. 
The main assumption of our method is that the \textit{intrinsic} colour-magnitude distribution at high Galactic latitude is conditional {only} on the absolute perpendicular distance above the Galactic plane $Z$ and the Galactic radial coordinate $R$. Moreover, stars are entered into the catalogue by satisfying a selection function which depends both on sky position and radial distance.

\subsubsection{Modelling the colour-magnitude distribution as a normalising flow}
To model the probability distribution of the catalogue colour-magnitude relations, we use conditional normalising flows \citep{winkler2019learning}, a neural-network-based conditional density estimator. Normalising flows originally emerged as generative models, but in the context of our work, their strength lies in their ability to provide density estimation, which, in theory, can be exact. {Normalising flows assume} that there is a function
\begin{equation}
    f_{\bm{\theta}}:\mathbb{R}^{\dim \boldsymbol{x}}\times \mathbb{R}^2 \rightarrow \mathbb{R}^{\dim \boldsymbol{x}} 
\end{equation}
which is an invertible bijection in the first argument that maps the data distribution conditional on $(R,Z)$ to a random variable,
$\bm{u}=f_{\bm{\theta}}(\bm{x},(R,Z))$, where $\bm{u} \sim \pi(\bm{u})$ is sampled from a standard multivariate normal and ${\bm{\theta}}$ are the function parameters to be learnt. The change of variables formula for probability distributions dictates that the probability distribution over $\bm{x}$ given $(R,Z)$ is given by \begin{equation}
        p_{\bm{\theta}}(\bm{x}|\,R,Z)=\pi(f_{\bm{\theta}}(\bm{x},(R,Z)))|\det ({\nabla_{\boldsymbol{x}} f_{\bm{\theta}}(\bm{x},(R,Z)){)}}|,
\end{equation}
where $\nabla_{\boldsymbol{c}} f_{\bm{\theta}}(\bm{c},(R,Z){)}$ is the Jacobian of the transformation. Learning the conditional distribution amounts to fitting for the parameters $\bm \theta$ of the bijection. In this paper, we use a Block Neural Autoregressive Flow (BNAF; \citealt{decao2019blockneuralautoregressiveflow}) to learn the desired distribution. A BNAF parametrises the bijection as a block lower triangular matrix with positive block diagonal weights and bijective activation functions, and is optimised for use in density evaluation due to having a non-analytic inverse. In this paper, we choose a hyperbolic tangent activation function. The BNAF architecture has the benefit of being a flexible distribution which is easy to compute, differentiable, and easily implemented with MCMC methods. We implement our model using  Google's \texttt{Jax} \citep{jax2018github} automatic differentiation package and train it using the Adam optimiser \citep{kingma2017adam} to find the parameters that minimise the loss function. For the flow hyperparameters $\boldsymbol{\theta}$, we choose a block dimension of $4$ and a layer depth of $3$, which gives $\dim \bm{\theta}=24{\times} (2{\times}2+\dim \boldsymbol{x})^2$, where $2$ corresponds to the dimension of the vector $(R, Z)$.
{We delay explicit discussion of the loss function, opting first to describe our full model.}

\subsection{Accounting for the Malmquist bias}
Each survey will have an associated faint limiting magnitude $m_{\rm faint}$ and a bright saturation limit $m_{\rm bright}$ such that objects only make it into the catalogue if the observed apparent magnitude in that passband satisfies $m_{\rm faint}>\hat m>m_{\rm bright}$. We constrain the parameter space allowed in our model {by} the saturation cuts, known as the Malmquist bias, based only on the leading absolute magnitude $M$ of the vector $\boldsymbol{x}$ and the distance $d$. Assuming that $m_{\rm faint}$ and $m_{\rm bright}$ are the limits in this passband, the absolute magnitudes of our model must satisfy the uniform density given by \begin{equation}
    I(M|\,\mu)=U(M|\,m_{\rm bright}-\mu,m_{\rm faint}-\mu).
\end{equation}
This density is multiplied by $p_{\bm{\theta}}(\bm{x}|\,R,Z)$ to truncate parameter space so that we can account for the possible absolute magnitudes that could make it into our catalogue given the distance. {This allows us to account for the Malmquist bias, the distance-dependent effects of the selection function.}

\subsection{Selection probability}
As we aim to learn the distribution, \begin{equation}
p_{\boldsymbol{\theta}}(\boldsymbol{x} |\,R, Z)\approx p^*(\boldsymbol{x} |\,R, Z)\end{equation} 
of zero-extinction stars, the probabilistic model must account for the selection effects of the data. In this paper, $S=1$ indicates the selection variable that the star would have been selected in the training sample if there were no photometric dust extinction. This selection variable is always satisfied in the training set, as we assume the training set consists of zero-extinction stars.
Let $\mathfrak{R}:=\{\boldsymbol q:\, \mathfrak{S}(\boldsymbol q)>0\}$ be the coverage set of data in terms of quantities $\boldsymbol{q}$ where $\mathfrak{S}$ is the selection function. In this paper, we assume that the selection function $\mathfrak{S}(\boldsymbol{ q})\approx \mathfrak{S}(\boldsymbol{\hat q})$ over the latent values of $\boldsymbol{q}$ is the same as over the observed values $ \boldsymbol{\hat q}$. 
After choosing a parent catalogue, {the process of} cross-matching with other surveys and making selection cuts to the data corresponds to generating {subsamples} of the parent catalogue. To this end, we use subsample selection function estimation methods {\citep{rix2021selection,castro2023estimating}} to estimate the probability of a source being included in our subsample of the parent catalogue for a given LOS. As mentioned {in the previous section}, we decouple the selection effects from the Malmquist bias and consider only sky-position-dependent selection functions.
The subsample method relies on the basic selection function of the parent catalogue being well understood. The selection function is a function $\mathfrak{S}_\mathcal{C}(\boldsymbol{q})$ that returns the probability of a source making it into the subsample. As in {\citealt{rix2021selection}}, this is given by
\begin{equation}
\mathfrak{S}_\mathcal{C}^{\mathrm{subsample}}(\boldsymbol{q}) \;=\; 
\mathfrak{S}_\mathcal{C}\bigl(\boldsymbol{q} \,\vert\, \boldsymbol{q}\in \text{parent}\bigr) \,
\cdot \,
\mathfrak{S}_\mathcal{C}^{\mathrm{parent}}(\boldsymbol{q}),
\label{eq:subsample_selection}
\end{equation}
where $\mathfrak{S}_\mathcal{C}^{\mathrm{parent}}(\boldsymbol{q})$ describes the probability that a source with attributes 
$q = \{\ell, b, M, \dots\}$ 
is included in the parent catalogue, and 
$\mathfrak{S}_\mathcal{C}\bigl(\boldsymbol{q} \,\vert\, \boldsymbol{q} \in \text{parent}\bigr)$ 
is the probability that a source is included in the subsample, given that it is already included in the parent catalogue.
In our generative model, we define $\boldsymbol{q}=(\boldsymbol{m}^{\rm  unext},  \hat l , \hat b)$. After adopting the appropriate selection function, we define the probability of the selection variable being true as \begin{equation}
    P(S=1|\, \boldsymbol{m}^{\rm unext},  \hat l , \hat b)=\mathfrak{S}(\boldsymbol{m}^{\rm unext},\hat l , \hat b),
\end{equation}
where $\boldsymbol{m}^{\rm unext}=\boldsymbol{M}^{\rm int}+5\log_{10}(d)-5$ in the forward model is the unextinguished apparent magnitude. That is, the photometry before the effects of dust extinction are applied in the forward model. The selection function depends on the unextinguished apparent magnitudes because we define the selection variable, $S$, as the selection variable associated with a star appearing in the original training sample.

\subsection{Galactic cylindrical coordinates in the forward model}
We adopt deterministic relations for the Galactic cylindrical coordinates from the assumption that $\hat l$ and $\hat b$ are known exactly. We will not have uniform coverage of $(R,Z)$ from the training sample, and we must define the set of possible $(R,Z)$ that our model has observed in the training phase. When inferring the posterior in a different LOS, utmost care must be taken to ensure that the inferred LOS remains within the allowable values of $(R,Z)$, and we will ensure that $(d|\sin(\hat{b})|\in Z_{\rm train})$ and $(R(d,\hat l,\hat b)\in R_{\rm train})$.

\subsection{Distance prior}

For the prior on distance, we adopt the Generalised Gamma distribution \citep{Bailer_Jones_distance}, known as the geometric prior therein,

\begin{equation}\label{bjp}p(d|\, \hat l,\hat b)=\begin{cases}
        \frac{\alpha}{\Gamma(\alpha^{-1}(\beta+1))} \frac{d^\beta}{L^{\beta+1}} \exp(-(\frac{d}{L})^{\alpha}) & \text{if } d >0 \, \rm pc\\
        0 & \text{if } d=0  \, \rm pc
    \end{cases}
\end{equation}
where the values of $\alpha, \beta$ and $L$ depend on the sky-position, $\hat l, \hat b$, and were {fit} to a three-dimensional model of our Galaxy. These values can be queried from the accompanying data products of \citealt{Bailer_Jones_distance}.

\subsection{Reddening prior and extinction law}\label{reddening_law}
We adopt an extinction law $\boldsymbol{A}=\boldsymbol{{R}}(E,C)E$ as a function of the interstellar reddening $E:=E(B-V)$ and an \textit{intrinsic} photometric colour $C$.
The extinction coefficients, $\boldsymbol{{R}}(E,C)$, of broadband filters depend on the underlying SED of the source and the amount of dust between the source and the observer \citep{Danielski_2018}, in particular when the passband is wide. Extinction laws are adopted either by assuming or inferring the relation, both of which are applicable within the scope of our model. In this paper, we choose to use the empirical colour- and reddening-dependent extinction law from {\citealt{2023ApJS..264...14Z}}, who {presented} the extinction coefficients $\boldsymbol{{R}}(E,C)$ as a function of either effective temperature or intrinsic colour and the reddening $E:=E(B-V)$ on the scale of {\citealt{SFD}}{. Therefore, the reddening values correspond to a single fixed value of $R_V\approx 3.1$. This allows us to identify the reddening $E$ with the extinction in the visible passband $A_V$.
The \citealt{2023ApJS..264...14Z} extinction coefficients are} a data-driven estimation of the extinction law for $0<E(B-V)<0.5 $ mag in terms of photometric colours. This range is ideal for the mid- to {low-extinction} regime at high Galactic latitudes. Moreover, the errors induced from adopting this extinction law are minimal for low extinction lines-of-sight \citep{high_gal_me}. The extinction law is mostly valid for stars with $4,000<T_{\rm eff}<10,000K$. At high Galactic latitudes, we will observe a significant number of stars with $T_{\rm eff}<4000$K, particularly from nearby faint stars. It is safe to assume this extinction law for these cases{,} as extinction is low in these regions, and the variation in the extinction law for a small change of reddening has a relatively small impact on the extinction coefficients.
We adopt a uniform prior on the reddening 
\begin{equation}
p(E)=\begin{cases}
        \frac{1}{0.5-E_{\rm min}} & \text{if } E \in (E_{\rm min},0.5)\text{ mag}\\
        0 & \text{otherwise} 
    \end{cases}
\end{equation}where we allow for negative $E_{\rm min}$ in our model to inspect the posterior without truncating it at zero. For the LOS used in this paper, we set $E_{\rm{min}}=-0.05$ mag. Moreover, our method defines its zero-point based on learning stars with \textit{negligible {extinction}} as opposed to zero {extinction}.

\subsection{Single star posterior distribution}
Assuming that we have trained the normalising flow (that is, $\boldsymbol{\theta}$ is determined) to model the probability density function of $\bm{x}$ given Galactic coordinates at high Galactic latitudes, we can infer the photometric reddening $E$ of a single star using Bayesian inference. For a given star, we observe $\hat {\bm \xi}$: a vector of the apparent magnitudes and parallax.

\subsubsection{Posterior distribution}
Given a star \textit{that has been included in the catalogue} and has an observed vector of parallax $\hat \omega$, Galactic latitude and longitude $ \hat l,\hat b$, and apparent magnitudes $\hat{\boldsymbol{m}}$. Let $\bm{\hat \xi}:=(\hat{\boldsymbol{m}},\hat \omega)$ denote the data vector and $\bm \Sigma$ denote a known covariance matrix corresponding to Gaussian measurement error on the photometry and $\sigma_\omega$ the Gaussian measurement error on the parallax. We wish to infer the dust reddening by calculating the posterior distribution $P(E|\,\hat{\bm{ \xi}},\hat l, \hat b, S=1)$ of reddening conditional on the observed data and assuming that the star made it into our catalogue. Throughout, $\boldsymbol{m}$ refers to the latent apparent magnitude constructed from $\boldsymbol{x}$ and $d$.
Let $\boldsymbol \Phi:=(R,Z)$ and $S$ be the binary variable indicating the selection function. The posterior distribution of reddening for a single star given the data $\bm{\hat \xi}$ is
\begin{equation}
    p(E|\,\bm{\hat \xi},\hat l, \hat b,S=1)= \int d d \,d\boldsymbol x \,p(E , \bm{x}, d|\,\bm{\hat \xi},\hat l, \hat b,S=1).
\end{equation}
Using Bayes' Theorem, the full unnormalised posterior factorises as
\begin{equation}\label{post}
\begin{aligned}
p(E , \bm{x}, d|\,\bm{\hat \xi},\hat l, \hat b,S=1) &\propto p(\bm{\hat m}|\, \boldsymbol{{x}},E,\mu)\\
&\times p{(\hat \omega} |\, d)p(E)\\&\times p_{\boldsymbol{\theta}}(\boldsymbol{{x}}|\,\boldsymbol \Phi(d, \hat l ,\hat b))I(M|\,\mu)\\
&\times P(S=1|\boldsymbol{m}^{\rm unext},\hat l, \hat b)p(d |\, \hat l, \hat b).
\end{aligned}
\end{equation}
{\subsubsection{Assumptions About the Inference Set}}
In our posterior formulation, we always assume that the underlying point has an intrinsic (unextinguished) colour-magnitude vector that is consistent with the selection function of the original training sample ($S=1$). We do not model the effect of stars entering the catalogue (passing the bright limit) because of the effects of dust extinction. To model this effect, we would need to describe $p(E|\,\bm{\hat \xi}, \hat l, \hat b ,S=0)$, the posterior distribution of the reddening given that the star has $S=0$, it only makes it into our inference set is \textit{because of dust extinction}. 
This is important to be aware of in the inference step. We must be careful to ensure that the observed point would be consistent with the selection function of the original training sample in the absence of dust extinction. We can query the integrated reddening value $E_{\rm max}$ from an independent dust map to generate an updated apparent magnitude cut $m_{\rm bright}^{\rm inference}=m_{\rm bright}+R_m E_{\rm max}$, where $R_m$ is the extinction coefficient in the leading passband, to flag whether the star could be pushed into the inference set because of the dust extinction.

\subsection{Training the {flow}: loss function definition}
Now that we have established the posterior for a single star, we can look to training the normalising flow by assuming that we have selected regions devoid of interstellar extinction.
We do not have samples from the true, noise-free target distribution $p^*(\bm{x},R,Z)$, so usual normalising flow training by maximising the log-likelihood of the parameters given the data directly from the flow does not apply. Rather, we must {consider} the marginal likelihood. The marginal likelihood is a density over the data that accounts for and marginalises over the latent variables, selection effects, and measurement error. It allows us to connect the normalising flow distribution to the observed data and learn the latent distribution. We train the normalising flow on a subsample of stars that we assume have zero extinction, thereby dropping the reddening term from the posterior model.
\subsubsection{Marginal {likelihood} for the full dataset}
For each star, we only have noisy estimates of the apparent magnitudes and parallax, which we denote by the vector 
\begin{equation}
    \xi^{(s)}=(\hat m_{1},...,\hat m_n,\hat \omega)^{(s)}, \quad D=\{\xi^{(s)}\}_{s=1}^N,
\end{equation}
where $D$ denotes the full dataset over all stars in the training sample. The total marginal likelihood of $\boldsymbol{\theta}$ is given by
\begin{equation}
    \mathcal{L}(\boldsymbol{\theta}) 
:= \prod _{s=1}^N L_s(\boldsymbol{\theta}),
\end{equation}
where \begin{equation}
    L_s(\boldsymbol{\theta})=p(\boldsymbol{\xi}_s\mid \hat l_s, \hat b_s,\boldsymbol{\theta}, S_s=1).
\end{equation}

\subsubsection{{Contrastive term}}
{
In general, normalising flows are known to be poor at identifying out-of-distribution data \citep{kirichenko2020normalizing}. This is problematic for using normalising flows for inferring the reddening of extinguished stars, as many highly extinguished stars will appear out-of-distribution relative to the zero-extinction training samples. To address this, we implement a contrastive approach to allow the normalising flow to learn to assign consistent low probability to extinguished stars.}

{We implement a contrastive component of the loss, that is, a term which assigns low probability density to points not in the training set. We consider}
\begin{equation}
    \mathcal{C}(\boldsymbol{\theta}) 
:= \prod _{s=1}^N C_s(\boldsymbol{\theta},E_s),
\end{equation}
where \begin{equation}
    C_s(\boldsymbol{\theta},E_s)=p(\boldsymbol{\xi}_s+(\boldsymbol{A}(E_s),0)\mid \hat l_s, \hat b_s,\boldsymbol{\theta}, S_s=1),
\end{equation}
{corresponds to the marginal likelihood of a datum if it had extinction $\boldsymbol{A}(E_s)$ added to each passband, where in training the reddening is sampled at random $E_s\sim U(0.05\, \rm{mag},0.5\, \rm{mag})$. Contrastive approaches for normalising flows must be implemented with care, as a trivial solution for negative samples is to let the log probability tend to negative infinity. Following \citealt{elsharkawy2025contrastive}, we ensure that the log-probability of the normalising flow is clipped at $-70$ for the negative samples.}
\subsubsection{Loss function}
{The loss function that we minimise is}
\begin{equation}
    {\mathcal{L}_{\rm FLOWER}(\boldsymbol{\theta}):=\lambda\log \mathcal{C}(\boldsymbol{\theta})-\log\mathcal{L}(\boldsymbol \theta),}
\end{equation} the total negative marginal log-likelihood of the model parameters given the data, which is the sum of the individual negative log-marginal likelihoods for a single star, indexed by $s=1,...,N$, {together with the contrastive term whose strength is controlled by $\lambda$. The hyperparameter $\lambda$ is increased slowly from zero during training until the marginal likelihood term decreases. The final value of $\lambda$ was $0.3$.}
\subsubsection{Marginal likelihood for a single star}
 The likelihood must be re-weighted according to the selection probability from the sky-position selection function \citep{2017arXiv170603856H} to account for the selection effects in the entire loss function. This results in the marginal likelihood for a single star, given by the following equation:
\begin{equation}
\begin{aligned}
p(\hat{\boldsymbol{\xi}}_s \mid \hat l_s, \hat b_s,\boldsymbol{\theta}, S_s=1) 
&= \frac{1}{
P(S_s=1 \mid  \hat l_s, \hat b_s,\boldsymbol{\theta})
}\\ & \quad \times
\displaystyle \int \! p(\hat{\boldsymbol{\xi}}_s \mid  {\bm{x}}_s, E_s=0,\mu_s)  \\ & \quad \times \,
p_{\boldsymbol{\theta}}({\bm{x}}_s \mid \boldsymbol{\Phi}(d_s, \hat l_s, \hat b_s))
 \\
& \quad \times
I(M_s|\,\mu_s)p(d_s \mid \hat{l}_s, \hat{b}_s) \,\\&\quad \times
P(S_s=1 \mid  \boldsymbol{m}^{\rm unext}_s, \hat l_s,  \hat b_s)\\&\quad
 \, d\boldsymbol{x}_s \, dd_s  ,
\end{aligned}
\end{equation}
where \begin{equation}
\begin{aligned}
P(S_s=1 \mid  \hat l_s,  \hat b_s,\boldsymbol{\theta}) &= \int 
p_{\boldsymbol{\theta}}({\bm{x}}_s \mid \boldsymbol{\Phi}(d_s, \hat l_s, \hat b_s))
 \\
& \quad \times
I(M_s|\,\mu_s)p(d_s \mid \hat{l}_s, \hat{b}_s) \,\\&\quad \times
{P(S_s=1 \mid  \boldsymbol{m}^{\rm unext}_s, \hat l_s,  \hat b_s)}\\&\quad
 \, d{\bm{x}}_s \, dd_s{,}
\end{aligned}
\end{equation}
{and \begin{equation}
p(\hat{\boldsymbol{\xi}}_s \mid  \bm{x}_s, E_s=0,\mu_s):=p(\bm{\hat m}_s|\, \boldsymbol{{x}}_s,0,\mu_s)
 p{(\hat \omega_s|\, d(\mu_s))}.\end{equation}}
We mitigate against selection effects from Malmquist bias through the multiplication of the normalising flow model with the truncation model $I(M_s|\,\mu_s)$.
Given a distance modulus $\mu$, the normalising flow is implemented so that the colour-magnitude vector $\boldsymbol{x}$ corresponds to unextinguished apparent magnitudes $\boldsymbol{m}^{\rm unext}$ that satisfy the Malmquist bias relations. That is, for each passband, the relation $m_{\rm faint}> m^{\rm unext}>m_{\rm bright}$ is ensured by forcing the density to be zero outside these regions. The adjustment of the likelihood accounts for the sky-position{-dependent} effects of the selection function.

\subsubsection{Implementing the loss function}
We can fit for the conditional distribution $p_{\bm{\theta}}(\bm{x}|\, R,Z)$ by {minimising the FLOWER loss function} over the training dataset of stars with respect to the flow parameters $\boldsymbol{\theta}$. {We} introduce importance sampling distributions in Appendix \ref{derivation} and derive a noisy estimate of the loss function that we implement in practice. The entire loss function is implemented by summing over all stars in the training sample. In this paper, cuts to the data were made so that the term from the selection function is nearly always very close to $1$ and the normalising constant has minimal impact on the correction of the density. We sample $K=32$ times for each of the $N$ stars from the importance sampling distributions and generate the $32N$ vectors to train the flow on. The terms of the sum are dominated by those which are generated by the importance sampling distributions. {We sample three negative samples for each star to be used in the contrastive component of the loss function.}

\subsection{Posterior inference}
For a single star, we fit the model to the data using the \texttt{NumPyro} \citep{phan2019composable} Python package and generate samples from the single star posterior (Equation \ref{post}) using the No U-Turns (NUTS, \citealt{hoffman2011nouturn}) sampler. We sample $10,000$ effectively independent samples after a $4000$ sample warm-up stage that is used to initialise the chain, and adapt the step size and the mass matrix. For a single star, it takes on average $20$ seconds to run the inference on a single CPU core.

\begin{figure}
    \centering
    \includegraphics[width=0.5\textwidth]{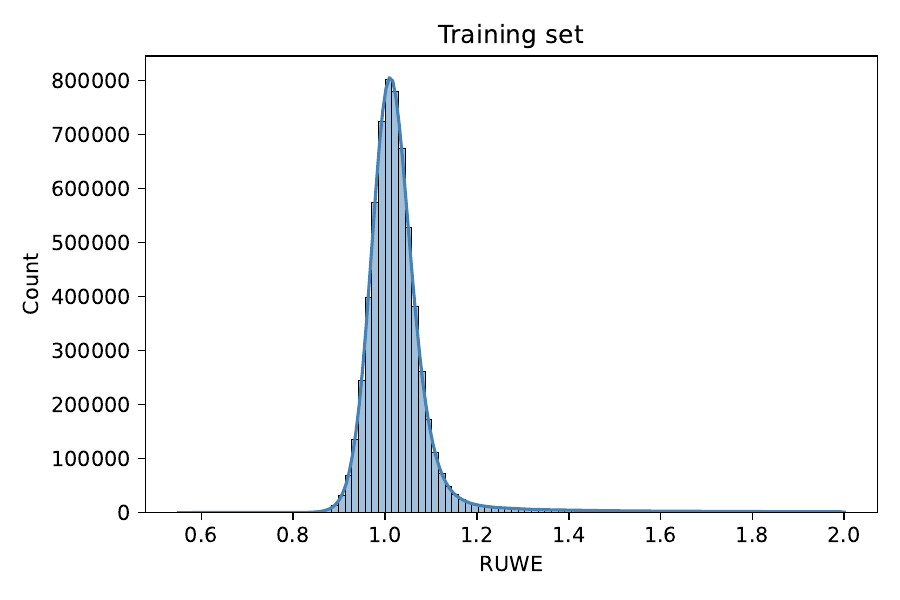}
        \caption{Distribution of Gaia's RUWE value for sources in the training set. In this paper, we wish to make a minimal number of cuts which could impact both the selection function and bias the extinction posterior in a manner that is difficult to quantify. The non-conservative cut of RUWE<2.0 used in this paper is reliable, as it leaves most of the sources in the training set within the literature-standard desired region of RUWE<1.4.}
    \label{fig:ruwe}
\end{figure}




\section{Data}\label{data_section}
We developed our model so that, in theory, any photometric survey that has reliable parallax or distance estimates can be used to train the normalising flow. The main assumption of our model is that we have reliably learned  $p_{\boldsymbol{\theta}}(\boldsymbol{{x}}|\,R,Z)$, the probability distribution of a zero-extinction colour-magnitude vector of stars at high Galactic latitudes, conditional on the Galactic cylindrical coordinates. 
There is a trade-off between high spatial resolution and wavelength coverage. In this paper, we illustrate the reddening posterior when we choose data with less wavelength coverage and infer reddening posteriors without near-infrared (NIR) data. It is well-known that the NIR data represent a significant increase in information in comparison with optical photometry for inferring dust extinction. We will show that even without NIR data, the increased spatial resolution will allow us to construct a tracer of the ISM that can struggle to accurately determine the density of the dust, but {will have an} increased spatial resolution.

\subsection{Gaia DR3}
The European Space Agency's (ESA) Gaia mission \citep{GAIA_MISSION} is a pioneering survey that aims to provide a precise three-dimensional chart of more than a billion stars throughout our Galaxy, mapping their distance, motions, luminosity, composition, and temperatures. The satellite consists of a broad $G$ band filter (330-1050 nm), {$BP$} (330-680 nm) and $RP$ (630-1050 nm) low-resolution fused-silica prisms and a Radial Velocity Spectrometer (RVS) instrument. In 2022, the Gaia Data Processing Consortium released the Gaia Data Release 3 \citep{GAIADR3}, which provides a full astrometric solution (position, parallax and proper motions) for about 1.46 billion sources, together with photometric apparent magnitudes $m_G$, $m_{BP}$ and $m_{RP}$ with a limiting magnitude of $m_G \approx 21$ and a saturation limit of $m_G \approx 5$ \citep{GAIADR3}.
We choose not to use the Gaia BP/RP spectra due to the difficulty in accounting for systematic error in calibrating the observed BP/RP spectra, and modelling this process is an active research area. However, we use the derived integrated passband magnitudes. We apply the parallax zero-point corrections as recommended in \citealt{zero-point}, which depend on position, magnitude, and colour. Similarly to {\citealt{Green_2021}}, we use the Gaia $G$ band as absolute magnitude $M_G$ in our model, and define all colours relative to $G$.
In the training step of our model, we exclude all Gaia sources which have {a} negative parallax. Moreover, in all datasets, we choose only sources with a $5$ parameter astrometric solution. Gaia provides a {Renormalised} Unit Weight Error (RUWE) that quantifies the astrometric deviations from single-source solutions. For models which depend on a single-star assumption, making effective cuts on RUWE is highly desirable due to its correlation with unresolved binaries \citep{Castro_Ginard_2024}. However, our model makes no underlying assumption about whether a star is binary or not, and RUWE impacts it through errors in the parallax determinations. Therefore, we make a {less} conservative cut than the standard of $\texttt{RUWE}<2.0$ for our training sample. To ensure high-quality sources, we require at least 9 visibility periods and 3 G{-}band observations.

\subsection{Pan-STARRS}
The Panoramic Survey Telescope and Rapid Response System 1 (Pan-STARRS, \citealt{PS1}), which is a 1.8 m optical and near-infrared telescope located in Hawaii, USA. Between 2010 and 2014, PS1 surveyed the sky north of declination $\delta>-30^\circ$ in five photometric passbands $m_g,m_r,m_i,m_z,m_y$ (grizy). The mean $5 \sigma$  point source limiting magnitudes in the stacked $3\pi$ Steradian Survey in grizy are (23.3, 23.2, 23.1, 22.3, 21.4) mag respectively. Saturation occurs around $m_g,m_r,m_i \approx 13.5$ mag, $m_z\approx 13.0$ mag, and $m_y \approx 12$ mag. The upper bound on the systematic uncertainty in the photometric calibration across the sky is 7-12 millimag, depending on the passband. In the training sample, we exclude all PS1 photometry flagged as unphotometric, saturated, or containing bad pixels.

\subsection{2MASS}
The Two Micron All Sky Survey (2MASS; \citealt{2MASS}). 2MASS is a ground-based, all-sky survey in three near-infrared passbands,  J (1.25 $\mu $ m), H (1.65 $\mu$ m), and $K_s$ (2.16 $\mu $ m) in the Vega system with a $1\sigma$ photometric uncertainty of <0.03 mag \citep{2mass06} and a limiting magnitude of $m_{K_s}\approx 14.3$ mag. 2MASS contains a total of 470 million sources, nearly 340 million sources of which are considered reliable. We only select cross-matched sources with the highest 2MASS photometric ('AAA') quality flag.

\begin{figure}
    \centering
    \includegraphics[width=0.5\textwidth]{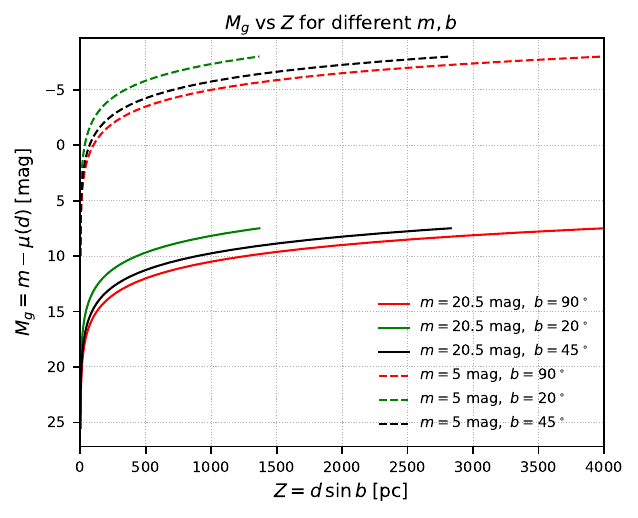}
        \caption{Illustrating the Malmquist bias: we display the absolute magnitude versus the Galactic height $Z$ for three different Galactic latitudes $b$. We trace out the line-of-sight for $0<d<4000$ pc and set the bounds between the saturation limit and the faint cut we impose on the apparent Gaia $G$ magnitude.}
    \label{fig:distance_z_mag}
\end{figure}

\subsection{Gaia selection function from subsamples}
We assume that Gaia represents the parent catalogue in our methods, and each of the other surveys and the cuts applied {corresponds} to generating a training or inference subsample of the Gaia catalogue. To this end, we use the selection function estimation from \citep{castro2023estimating} to estimate the probability of a source being included in our subsample of the Gaia catalogue.

\section{Learning zero-extinction colour-magnitude relations}\label{training_regions}
We illustrate our method by training on regions at high-Galactic latitude with minimal extinction, and infer the {reddening} of stars towards a different LOS. The sample of stars used for training the model depends on the experiment one wishes to {carry out}. In our implementation, we consider a more general distribution that learns the colour-magnitude distribution of stars along LOS with $E(B-V)_{\rm{Lenz}}<0.018$ mag from the {\citealt{Lenz_2017}}. For the data used in this paper, the errors on the extinction law and the photometry make it unlikely that we can reliably adopt an external extinction law and infer the reddening for stars with $E(B-V)_{\rm{Lenz}}<0.018$, so this is a reasonable place to start. With more accurate data, or if one wishes to do experiments on the extinction law in low-extinction regimes, then they may wish to choose a training set of stars with lower $E(B-V)_{\rm{Lenz}}$.

\begin{table}
    \centering
    
    \label{tab:photometry_summary}
    \begin{tabular}{|l|c|}
        \hline
        
        \textbf{Survey} & \textbf{Star Count} \\
        \hline
        Gaia + PS1 & $7187588$ \\
        + 2MASS & $3474886$ \\
        \hline
        \textbf{Quantity} & \textbf{Median Value} \\
        \hline
        Parallax $\omega$ error  & $0.07138$ mas \\
        Gaia $G$ error & $0.00060$ mag\\
        Gaia $G_{\rm BP}$ error & $0.01054$ mag\\
        Gaia $G_{\rm RP}$ error & $0.00307$ mag\\
        PS1 $g_{\rm PSF}$ error & $0.00725$ mag\\
        PS1 $r_{\rm PSF}$ error & $0.00581$ mag\\
        PS1 $i_{\rm PSF}$ error & $0.00400$ mag\\
        PS1 $z_{\rm PSF}$ error & $0.00433$ mag\\
        PS1 $y_{\rm PSF}$ error & $0.00569$ mag\\
        2MASS $J$ error & $0.03300$ mag\\
        2MASS $H$ error & $0.04000$ mag\\
        2MASS $K_s$ error & $0.05100$ mag\\
        \hline
    \end{tabular}
    \caption{Star count and median measurement error values from the training set $E(B-V)_{\rm Lenz}\leq 0.018$ mag. We highlight that the 2MASS cross-match, which provides valuable information for inferring dust reddening, comes at the cost of a significant cut to the total star count.}
\end{table}
\begin{figure*}
    \centering
    \includegraphics[width=\textwidth]{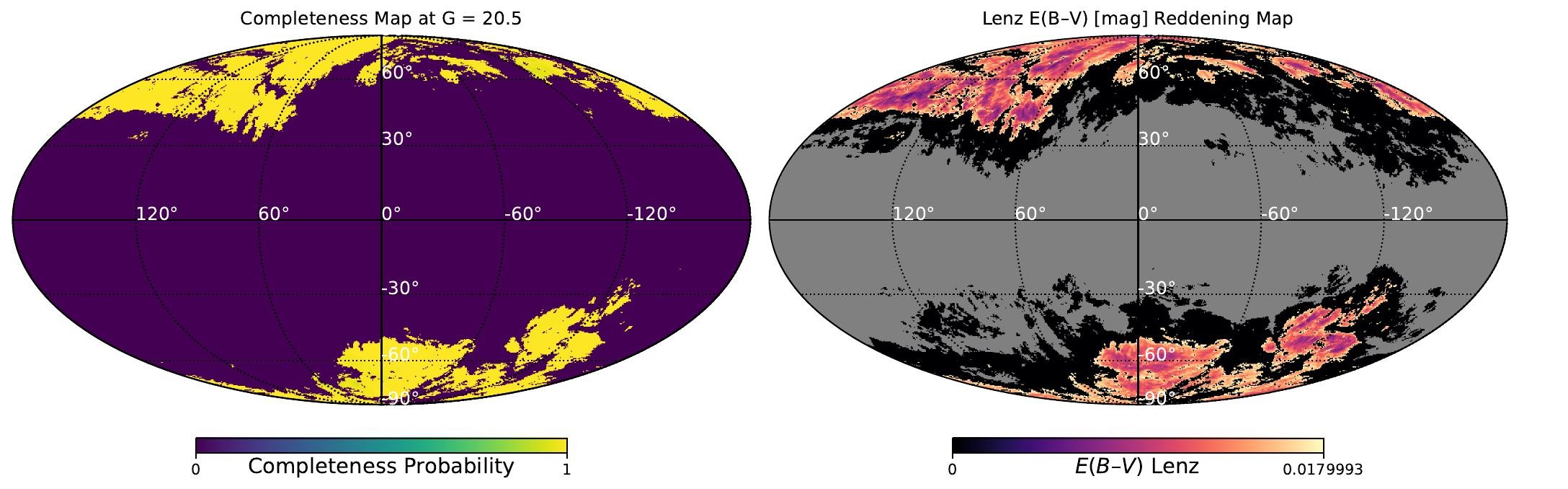}
        \caption{The training sample regions in Galactic coordinates. The target regions were chosen by querying the \citealt{Lenz_2017} dust map for zero-extinction regions. The plot on the right displays the Lenz $E(B-V)$ values subject to the cut of $E(B-V)_{\rm Lenz}<0.018$ mag, with the black hue indicating regions of Lenz map coverage that were not chosen for the training set. The plot on the left shows the completeness map of Gaia using the Gaia unlimited package at $G<20.5$ for the training set. This value was chosen so that there is nearly uniform coverage at this magnitude for our training lines of sight.}
    \label{fig:training_region}
\end{figure*}

\subsection{Training region: $E(B-V)_{\rm{Lenz}}<0.018$ mag}
We query Gaia DR3 for all stars with $E(B-V)_{\rm{Lenz}}<0.018$ mag from the \citealt{Lenz_2017} dust map and generate two training datasets by cross-matching with PS1 and 2MASS. Furthermore, we apply the cuts as described in Section \ref{data_section}. We display the regions in Galactic coordinates in the right plot of Figure \ref{fig:training_region}. We intentionally make very few cuts to the underlying training to inspect the reddening posteriors under these assumptions. If one can perform informative cuts {which will not impact the selection of stars for reddening posteriors}, {these} can be adopted {but may introduce further bias into the sample. Provided these biases can be quantified, then they are} valid and recommended.
\subsubsection{Gaia {selection function} from {subsamples}}
We display a plot of the selection function probability for our training sample in the left plot of Figure \ref{fig:training_region_logp}.

\subsubsection{Data Description}
After the recommended cuts, we have a sample size of $N={7,\!187,\!588}$ stars that pass the quality cuts for this region using only Gaia and PS1. When we train our model including 2MASS data, we lose around $50\%$ of the stars in the training region. The quoted errors of the photometry and the parallax are low for this large sample (see Table \ref{tab:photometry_summary}). We display the Gaia colour-magnitude diagram in Figure \ref{fig:training_region_hr} for the sample of data using only Gaia and PS1 data. We can see that the underlying training data is relatively agnostic to the stellar type that we have observed. This is intentional, as we do not wish to make any harsh cuts that could bias our sample in an undesirable manner and make the selection function more difficult to quantify when inferring reddening. Moreover, we make a relatively high cut to the RUWE of $RUWE<2.0$. However, we see that the bulk of the RUWE distribution for our sample falls in an agreeable region of $RUWE<1.4$ (Figure \ref{fig:ruwe}). Our goal is to learn the catalogue photometry of sources, and making specific cuts to the training sample may be difficult to propagate into the selection function both in training and in inference, potentially biasing the reddening inference at a later stage.

\subsubsection{Z-R Coverage}
We must have good coverage of Galactic $Z-R$ in the training sample if we wish to infer extinction along a different LOS. The target LOS must lie entirely within the region covered by the training sample within $2.5$ kpc. In Figure \ref{fig:inference_pasiphae}, we show the coverage of the training sample. 

\subsection{Training the normalising Flow}
We train two separate flows, one using 2MASS data and the other without. Masking can be adopted if one wishes to use our method on partial data, but we ignore this in this paper. In the inference step, the correct flow will be selected automatically depending on the inferred data. In what follows, we will refer to a single normalising flow.
We train the flow on the sample of stars using the {FLOWER} loss function with a learning rate of $10^{-3}$ and {a} L2 regularisation with weight decay of $10^{-4}$. We choose a batch size of $2^{13}$ and train for a maximum number of 1000 epochs or until the validation sample performance has a maximum patience of 50 epochs. That is, the validation performance plateaus for 50 epochs. {We slowly increased the contrastive hyperparameter from $0$ to a value of $0.3$ during training.}
After training, we evaluate the performance of the fit and find that the normalising flow can faithfully recover the target distribution. In Figure \ref{fig:training_region_hr} we display an {HR} diagram with the hue indicating the log-probability density of the normalising flow evaluated on the data.

\subsection{Low probability sources on the edge of PS1 coverage}
\begin{figure*}
    \centering
    \includegraphics[width=\textwidth]{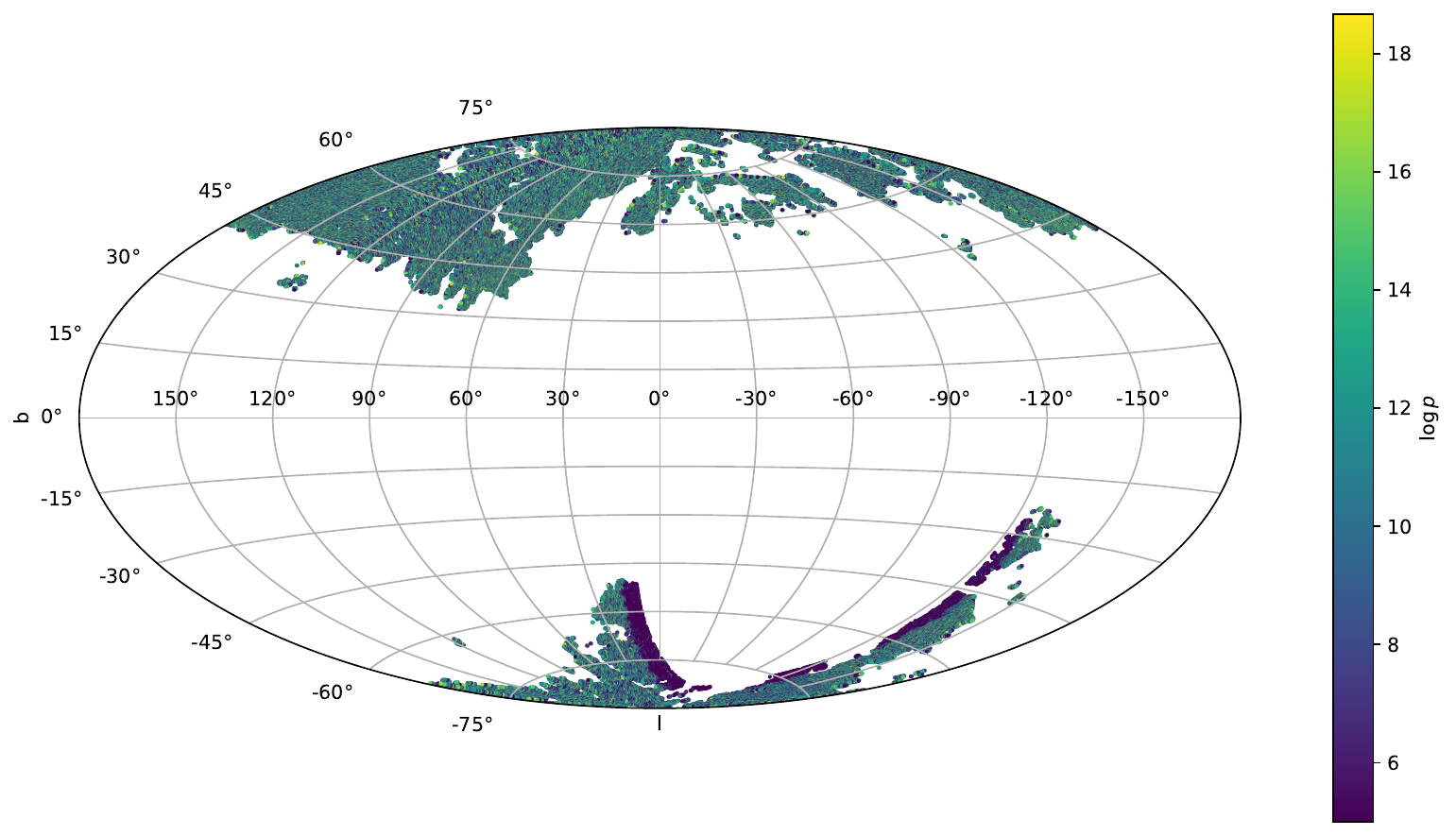}
        \caption{The log probability density from the normalising flow evaluated on the training region with hue minimum of 6. The consistent well-mixing of low and high probability regions across the sky position is expected, as we should expect no global correlation. However, in this plot we can see a significant structure at the boundary of PS1's coverage region (declination $-30^\circ$). Upon investigating this data, we find that the Gaia recommended cross-match is providing an incorrect match even after all recommended cross-match cuts are made. This is promising in the context of learning the true colour-magnitude relations, as in the PS1-$G$ colour space, these points are showing up in low probability regions relative to the entire training sample and can highlight inconsistencies in the data.}
    \label{fig:training_region_logp}
\end{figure*}

From the {normalising flow} log probability density map shown in Figure \ref{fig:training_region_logp}, which were generated by evaluating the flow directly on the noisy estimates of $\boldsymbol{x}$, $Z$, and $R$ calculated directly from the data, we see that there is certain spatial correlation visible with respect to the log{-}probability density. We find {that significant} low-probability structures {can be found} at the edge of the PS1 coverage (declination near $-30^\circ$). Further investigation shows that this discrepancy occurs due to the colour-magnitude relation in PS1 colours relative to Gaia $G$. We hypothesise that the Gaia cross-match is failing in this region due to the PS1$-G$ colours appearing in a far out-of-distribution part of colour-colour space. However, we stress that this has a negligible impact on the learning of the distribution provided that the errors are not consistent for all $Z,R$ covering these regions. In other words, our model can identify these points as outliers and assign low probability to them. This indicates that the model is robust to outliers on a large population scale and can highlight inherent errors in the data in a self-supervised manner. Removing these points and reiterating the training without them provides an identical distribution.

\begin{figure*}
    \centering
    \includegraphics[width=\textwidth]{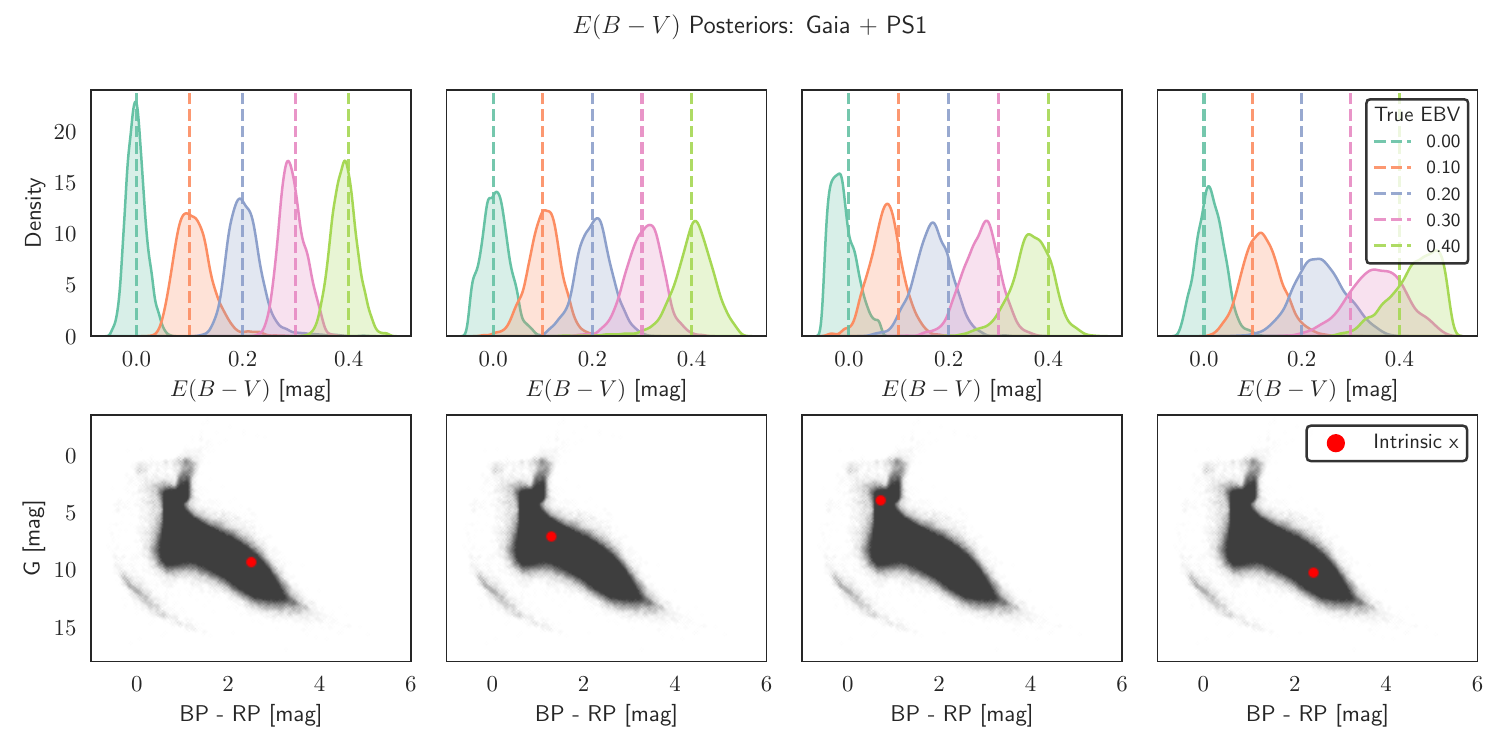}
        \caption{Posterior plots of photometric reddening for the synthetic validation set using only Gaia and PS1 data. The top column shows the reddening posterior for different amounts of added synthetic reddening $E_{\rm syn}=0.00,0.1,0.2,0.3,0.4$, where the synthetic reddening is added to the photometry from the low reddening stars validation set that was used in the training regime. In the bottom row, we use a red dot to indicate the intrinsic colour-magnitude point used to add photometric reddening to. As expected from the literature, we find varying posterior geometries for different parts of the colour-magnitude diagram. This occurs mainly due to the degeneracy between the extinction law and the colour-magnitude diagram being more severe at certain parts of the colour-magnitude diagram. Due to the small negative mass allowed by our reddening prior, if the standard deviation of the posterior is greater than 0.05, then the zero reddening posterior will have a slightly positively biased posterior mean.}
    \label{fig:posterior_validation_points}
\end{figure*}
\begin{figure}
    \centering
    \includegraphics[width=0.5\textwidth]{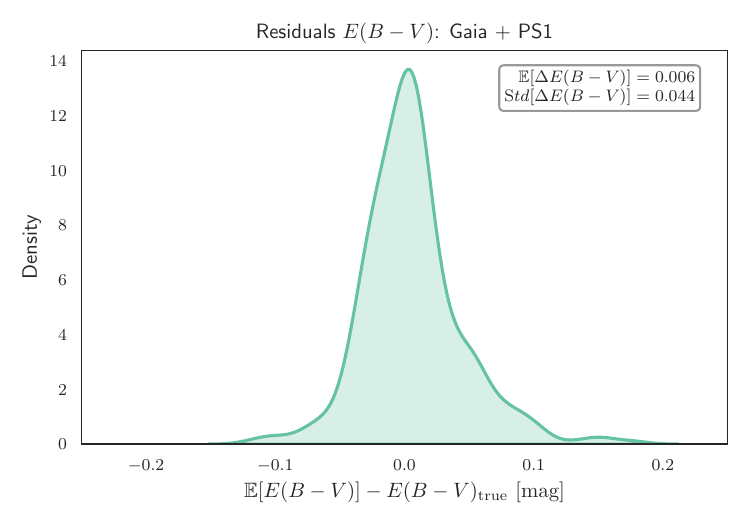}
        \caption{The posterior mean reddening residual against the synthetic truth for the synthetic reddening experiment. We find that the residuals are very reasonable. More importantly, we find consistent error bars across the colour-magnitude diagram so that points which cannot recover the true value precisely as a point estimate are accurate in that their error bars are wide. {The small positive biases that occur are expected as certain stars in the training set will appear as relatively low probability. Without prior distributions on reddening, the posterior probability may be higher than the value of 0. However, the width of the posterior will inflate to account for the uncertainty.}}
    \label{fig:ebv_residuals}
\end{figure}

\begin{figure}
    \centering
    \includegraphics[width=0.5\textwidth]{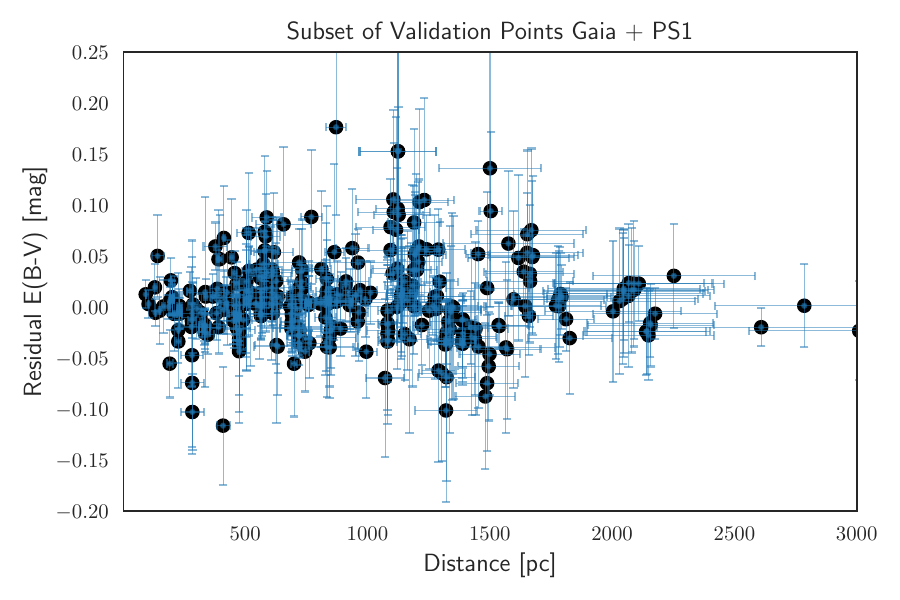}
        \caption{The posterior mean reddening residual from the synthetic true value versus the posterior mean distance for a subsample of the validation set using only Gaia and PS1 data. We find that the posterior can reliably recover the truth within the probability of the posterior. Any widening of the posterior will still give a significant probability for the truth. The discrepant points occur when the intrinsic points lie almost perfectly along the degeneracy of reddening with the photometric colour relations.}
    \label{fig:ebv_distance_validation_ps1}
\end{figure}

\begin{figure*}
    \centering
    \includegraphics[width=\textwidth]{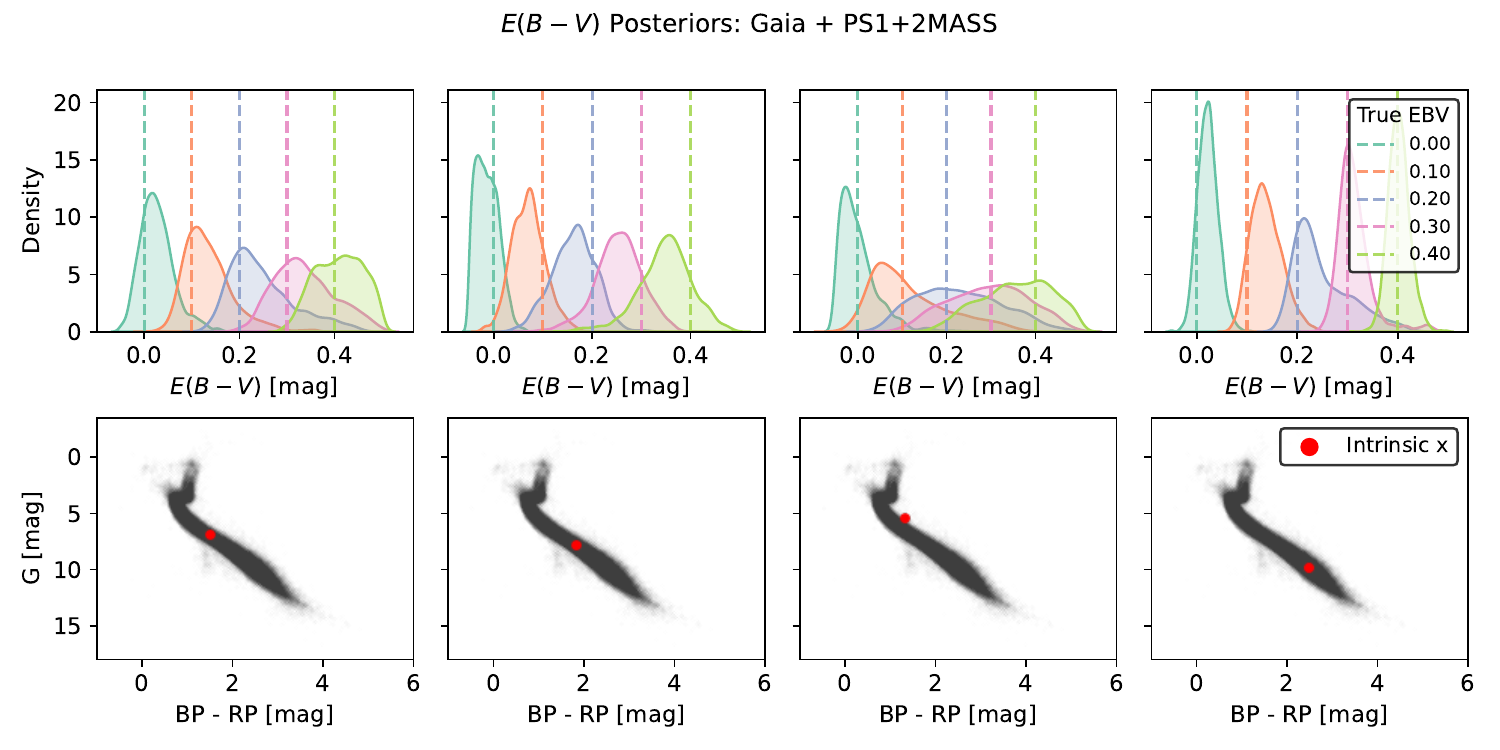}
        \caption{Posterior plots of photometric reddening for the synthetic validation set using Gaia, PS1, \textbf{and 2MASS} data. The top column shows the reddening posterior for different amounts of added synthetic reddening $E_{\rm syn}=0.00,0.1,0.2,0.3,0.4$, where the synthetic reddening is added to the photometry from the low reddening stars validation set that was used in the training regime. In the bottom row, we use a red dot to indicate the intrinsic colour-magnitude point used to add photometric reddening to. As expected from the literature, we find varying posterior geometries for different parts of the colour-magnitude diagram. This occurs mainly due to the degeneracy between the extinction law and the colour-magnitude diagram being more severe at certain parts of the colour-magnitude diagram. Due to the small negative mass allowed by our reddening prior, if the standard deviation of the posterior is greater than 0.05, then the zero reddening posterior will have a slightly positively biased posterior mean.}
    \label{fig:posterior_validation_points_2mass}
\end{figure*}
\begin{figure}
    \centering
    \includegraphics[width=0.5\textwidth]{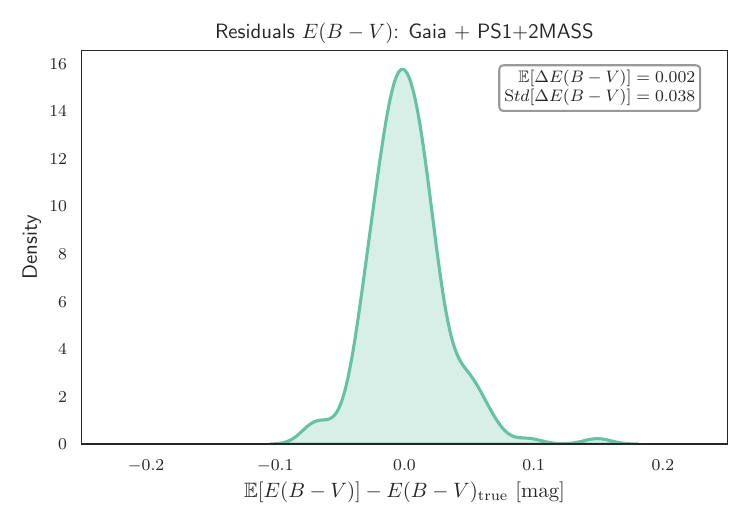}
        \caption{The posterior mean reddening residual against the synthetic truth for the synthetic reddening experiment using Gaia, PS1, and 2MASS. We find that the residuals are very reasonable. More importantly, we find consistent error bars across the colour-magnitude diagram so that points which cannot recover the true value precisely as a point estimate are accurate in that their error bars are wide. {The small positive biases that occur are expected as certain stars in the training set will appear as relatively low probability. Without prior distributions on reddening, the posterior probability may be higher than the value of 0. However, the width of the posterior will inflate to account for the uncertainty.}}
    \label{fig:ebv_residuals_2mass}
\end{figure}

\begin{figure}
    \centering
    \includegraphics[width=0.5\textwidth]{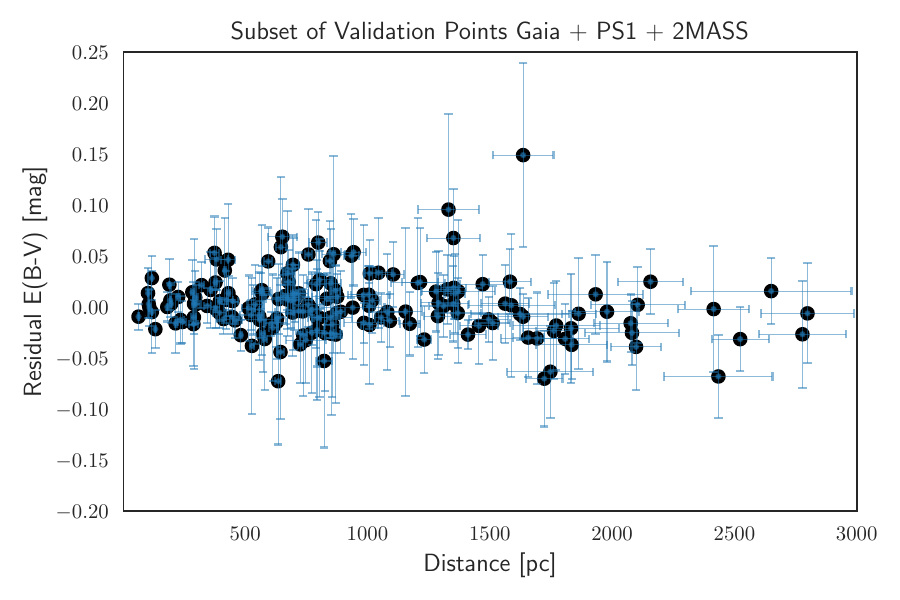}
        \caption{The posterior mean reddening residual from the synthetic true value versus the posterior mean distance for a subsample of the validation set using Gaia, PS1, and 2MASS data. We find that the posterior can reliably recover the truth within the probability of the posterior. Any widening of the posterior will still give a significant probability for the truth. The discrepant points occur when the intrinsic points lie almost perfectly along the degeneracy of reddening with the photometric colour relations.}
    \label{fig:ebv_distance_validation_2mass}
\end{figure}
\section{Reddening posterior inference}\label{inference_results}

\subsection{Synthetic reddening validation}
We validate our method's ability to recover reddening posteriors under the hypothesis of the model: that the observable properties of stars have been learnt correctly, conditional on Galactic Z and R. We found that we can recover reliable reddening posteriors where the true value obtains significant posterior mass. At certain points of the colour-magnitude diagram, there are biases in the individual reddening posteriors, but the width of the posterior distribution scales correctly to account for the true data-generating process. Our model can recover reliable posteriors and informative error bars in many but not all areas of the colour-magnitude diagram.

\subsubsection{Generating data with synthetic extinction}
To validate the accuracy of our model in recovering the reddening posterior, we take the validation set from the training split and synthetically add extinction to the photometry to generate a realistic dataset for which we infer the reddening. This dataset has been generated from a broad range of samples that represent a reliable subsample of the training set. 
Explicitly, we use the extinction law from {\citealt{Zhang_2022}} and generate synthetic $E$ values using the Gaia $BP-RP$ colour from the zero-extinction validation set. We assume that the distance is known exactly as the median posterior estimate from {\citealt{Bailer_Jones_distance}} and sample apparent magnitudes and parallaxes using the likelihood $N(\bm{\hat m}|\, \boldsymbol{\Lambda}^{-1}\boldsymbol{x}+\boldsymbol{{R}}(E,C)E + \boldsymbol{1}\mu(d), \boldsymbol{\Sigma})$ and inverting the distance. This generates a dataset with which we evaluate the performance of recovering the true synthetic reddening.
\subsubsection{Reddening validation}
We find that we can reliably recover the true synthetic reddening value under the given posterior distribution at varying accuracy across the colour-magnitude diagram. {Importantly, our model is well-calibrated, and} the true parameter is contained within the bulk of the posterior distribution. We evaluate the performance of the model with and without 2MASS data.
In Figure \ref{fig:posterior_validation_points}, we show the reddening posterior distribution for the point with intrinsic magnitude given by the red dots on the bottom row for different values of simulated extinction when using only Gaia and PS1 photometric data. From this subset of stars, we can see that the posterior distribution will be significantly different depending on the location of the point on the HR diagram. This is not surprising, as the degeneracy between the extinction law and intrinsic colours is more tightly coupled at brighter absolute magnitudes \citep{high_gal_me}. {We} can compare the reddening posteriors using Gaia, PS1, and 2MASS to those from Figure \ref{fig:posterior_validation_points_2mass}, where we display example posteriors for different points of the validation set. 
Furthermore, we display the residuals between the mean posterior reddening and the true synthetic reddening for Gaia+PS1, and Gaia+PS1+2MASS in {the left plots of} Figures \ref{fig:ebv_residuals} and \ref{fig:ebv_residuals_2mass}, respectively. In both cases, we find very reasonable reddening values. For the Gaia+PS1 set, the errors amount to a ${0.006}$ mag residual and an average posterior standard deviation across the samples of ${0.044}$ mag. For the Gaia+PS1+2MASS set, the errors amount to a ${0.002}$ mag residual and an average posterior standard deviation across the samples of ${0.038}$ mag.

For a sample of stars in the validation set, we plot the posterior mean reddening versus the posterior mean distances in Figures \ref{fig:ebv_distance_validation_ps1} and \ref{fig:ebv_distance_validation_2mass} with and without 2MASS, respectively. From these plots, it is evident that certain points of the colour-magnitude diagram can appear systematically different as point estimates, but their standard deviations indicate that the posterior width correctly inflates to account for the uncertainty in the estimates of these reddenings.

{The uncertainties on the extinction posterior reliably inflate to account for the discrepancy between the posterior samples and the truth. Some of the point estimates from the validation set can be biased slightly, which is evident from the fringes of the residual distributions. In a purely analytic, synthetic method for inferring the reddening, this would be concerning. However, our model is trained on data where the training set has varying probabilities for different stars, and some of these genuinely zero-extinction stars are relatively unlikely to be found at high Galactic latitudes. Without a strong reddening prior, the model assigns a higher probability to certain zero-extinction stars having reddening than their true position in colour-magnitude space. However, our model will inflate the error bars to reliably account for that, and it is uncertainty-aware on reddening, and when we cannot reliably recover the reddening truth, it will properly inflate the uncertainty. We can see an example of this case in the third column of Figure \ref{fig:posterior_validation_points_2mass}.}

\subsection{Two clouds at mid Galactic latitude}

We analyse two circular regions of radius $0.16^\circ$ region centred on $(l,b)=(104.08^\circ,22.31^\circ)$, known as the two-cloud region, and $(l,b)=(103.9^\circ,21.97^\circ)$, known as the one-cloud region. These regions have been studied in \citealt{Panopoulou_2019} by the PASIPHAE team \citep{tassis2018pasiphae}, where they show that the polarisation signal is a tracer of two large clouds along the LOS (each at varying distances), making them suitable as calibration test regions. They estimate the distance to the first cloud to be 346-393 pc and the second cloud to be 1250 - 2140 pc. This work was followed up with similar results in \citealt{2024A&A...684A.162P}. In this paper, we investigate these regions to see if the individual reddening posteriors give evidence to support the dust cloud distribution along the LOS. 

The first region we refer to as the two-cloud region, due to there being a clear signal of two clouds along the LOS using both polarisation and HI emission as a tracer for the clouds. The second region we call the one-cloud region, as the signal is dominated by the nearer cloud. Figure 1 in \citealt{Panopoulou_2019} shows the HI emission maps of the region, where the two-cloud region corresponds to the black circle, and the one-cloud region corresponds to the red circle therein.

We chose this region as it has been well-studied using independent polarisation measurements to constrain the dust cloud. Moreover, it has good coverage in the $Z-R$ plane of the training set out to $3.5$ kpc (Figure \ref{fig:inference_pasiphae}). Importantly, the second cloud in the two-cloud region consists of fine structural variation that is to be expected as typical of high Galactic latitude regions. In particular, the second cloud in the two-cloud region has an expected reddening differential $E(B-V)>0.02$ mag \citep{Panopoulou_2019} which should only be detectable on average if we have sufficient star counts.

\subsubsection{Posterior inference of reddening}
We query the data for the two regions using the cuts described in Section \ref{data_section}. The $Z-R$ coverage of these LOS is agreeable with respect to the training sample of stars (Figure \ref{fig:inference_pasiphae}) and therefore we expect to be able to recover reliable posterior inference of the reddening and distance for each star. In this paper, we do not concern ourselves with building reliable dust map images; instead, we focus on recovering the star-by-star posterior and leaving more sophisticated LOS inversions for future work. In general, our naive approach averaged over many LOS will have undesirable features \citep{edenhofer23} due to the difficulties in resolving distance along the LOS. However, we find that our posterior point estimates provide a \textit{tracer} of the dust cloud as opposed to accurately resolving the zero-point adjusted reddening as a function of distance.

We find that the LOS posterior mean extinction versus the posterior mean distance shows evidence of the cloud structure in both the Gaia+PS1+2MASS (Figure \ref{fig:ebv_distance_both_2mass}) and Gaia+PS1 (Figure \ref{fig:ebv_distance_both}) cases, with a better constraint on the overall reddening of the dust cloud in the former and a better estimate of the cloud's distance in the latter due to increased star count. We are not trying to use our point estimates to exactly constrain the reddening of each of the dust map components, due to the complex posteriors, which will be systematically different at different points of the colour-magnitude diagram. However, the posterior mean reddening along the LOS appears to be a tracer for the two clouds along the LOS. In both cases, we find evidence of the cloud structure of each LOS as indicative of the results of \citealt{Panopoulou_2019}.

In these plots, we see evidence of the trade-off between wavelength coverage and angular resolution. Without the 2MASS data, we have a far higher star count and, therefore, {higher spatial resolution} along the LOS. However, we have significantly more scatter and many stars appearing as low extinction stars, perhaps correctly, but in part due to the lack of NIR data. With 2MASS, the number of low reddening stars is lower, as the data has more wavelength coverage with less of a degeneracy between fitting for the intrinsic colours and the extinction. 

Our error bars are consistent along these LOS, and the points which show up as low reddening are there because they are buried in the middle of the main sequence in such a manner that there is no evidence of reddening without prior assumptions on the structure of the ISM. These assumptions are, therefore, necessary to construct consistent reddening estimates along the LOS, the most common of which is to assume that extinction is non-decreasing as a function of distance for a specified angular resolution. A priori, there is no guarantee that a fine, fractal geometry that the dust cloud may exhibit is compatible with these assumptions. In this paper, we show the reddening estimates without these assumptions.

\subsubsection{Comparison with existing dust maps and reddening estimates}
We compare our reddening profile results for the two circular regions to the results from the bayestar \citep{greenmap} and \citealt{Lallement} photometric dust maps, together with the spectro-photometric reddening estimates from \citealt{mlgreen}. The data is queried using the \texttt{dustmaps} \citep{2018JOSS....3..695M} python package and the data products from \citealt{Lallement} and \citealt{mlgreen}. For the bayestar map, we query \texttt{dustmaps} on a grid finer than the resolution of each of the maps within the circle of radius $0.16^\circ$. This is so that we can collect LOS profiles of extinction that each map generates in this region, as bayestar was generated on a resolution finer than $0.16^\circ$.

In Figure \ref{fig:bayestar_ebv}, we display the \citealt{greenmap} posterior mean reddening LOS profile, $E(B-V)_{\rm bayestar}$ versus the distance bins for the two-cloud (left) and one-cloud (right) regions. Their posteriors assume a certain spatial structure of the dust clouds. To generate these plots, we take 5 different lines-of-sight within the $0.16 ^\circ$ target region and overlay the posterior mean reddening from bayestar. Interestingly, their posterior only shows evidence of a two-cloud geometry in the \textbf{one-cloud} region, which is inconsistent with both the polarisation signal and HI data from \citealt{Panopoulou_2019}. Therefore, we conclude that bayestar struggles to resolve the fine variation on the scales of the target region. We hypothesise that this is due to the spatial smoothing effects of bayestar, which has invoked the structure of the region on a larger scale \citep{2024A&A...684A.162P} and smoothed over the finer variation on the quoted angular resolution. {We display the reddening profile for these regions from \citealt{Lallement} in Figure \ref{fig:lallement}, where their dust map only reaches 1 kpc at these latitudes. The first cloud is resolved along both LOS.}

We display the results in Figure \ref{fig:ebv_distance_green_zhang} for the reddening estimates from \citealt{mlgreen}. Their model was trained using Gaia BP/RP spectra, stellar parameter estimates from LAMOST, and \citealt{mlgreen} reddening values. In this plot, we display the \citealt{mlgreen} mean reddening, $E(B-V)_Z$, versus the mean distance for the two-cloud (left) and one-cloud (right) regions. {We display these for completeness, but note that their method is not entirely compatible with ours due to using a higher resolution dataset. The other, photometric dust maps, are a more faithful comparison to our model. However, they also encode prior knowledge about the dust structure into their published values.}

In both external dust extinction maps, we see a low extinction LOS where the main structure does not agree entirely with the results of the independent study using polarisation \citep{Panopoulou_2019}.

 \begin{figure*}
    \centering
    \includegraphics[width=\textwidth]{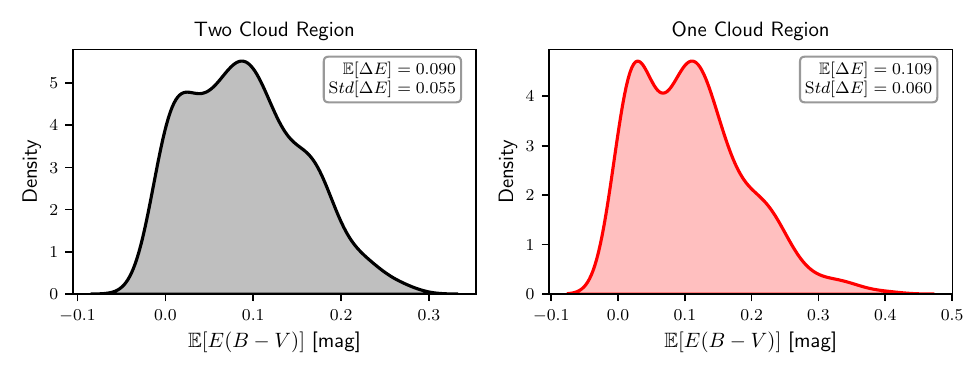}
        \caption{Distribution of FLOWER posterior mean reddenings across the sample of stars for the two cloud (left plot) and one cloud region (right plot) using only the Gaia and PS1 data. The overall zero point of our method should be different from other methods for estimating the reddening due to the difference in the forward model and the definition of our point estimates and zero point.  }
    \label{fig:both_region_ebv}
\end{figure*}

 \begin{figure*}
    \centering
    \includegraphics[width=\textwidth]{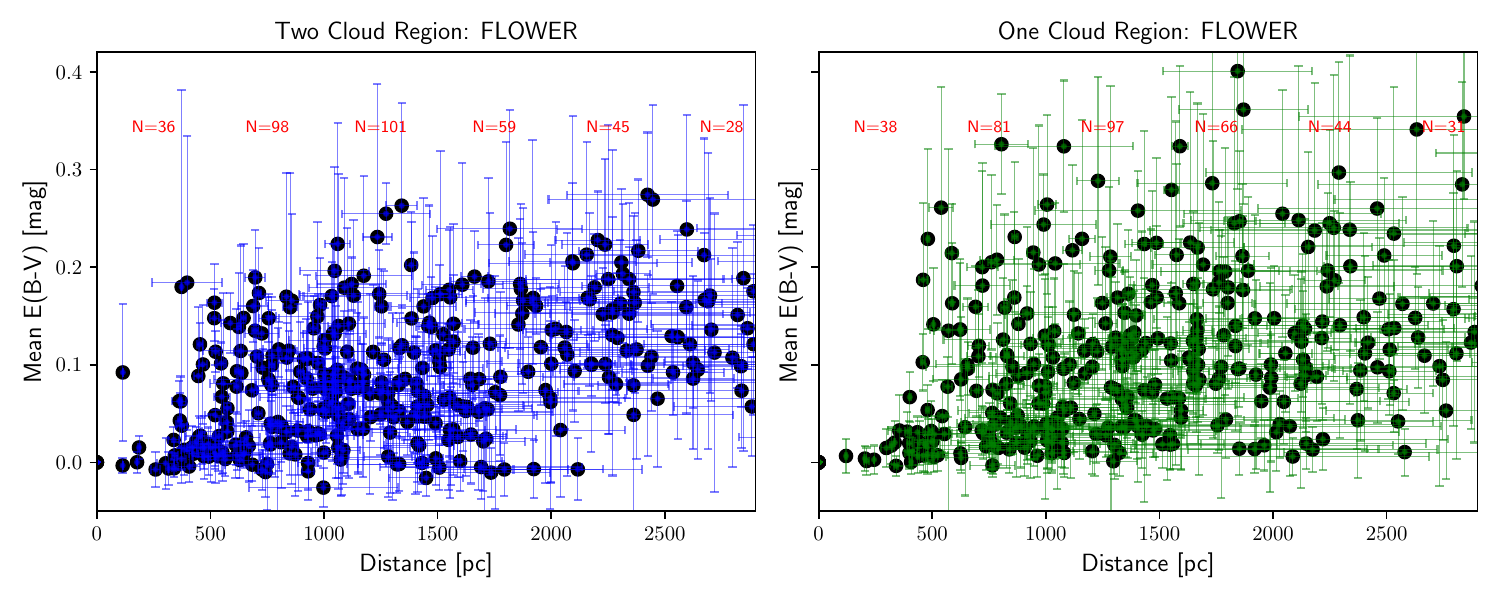}
        \caption{Gaia+PS1: The FLOWER posterior mean reddening versus the posterior mean distance for the two-cloud (left) and one-cloud (right) regions. It is important to note that our zero point was adopted from the negligible extinction training sample and should not agree in overall magnitude with other methods exactly. However, we find that our posterior mean reddening is indicative of the structure of the dust clouds along the LOS. In the two cloud region plot on the left, we can see that our point estimates give evidence for the first and second dust clouds along the LOS. Other points appear to have low reddening under the point estimates from our methods. Some of these points are appearing around the white dwarf binary part of the colour-magnitude diagram, and the best evidence for stars in this region is to assign low reddening. Other points appear right along the degeneracy and in the centre of the colour-magnitude diagram, where there is no evidence that the point should have higher reddening without making further spatial assumptions about the dust reddening relations as a function of distance. The numbers in red $N=j$ indicate the number of stars in the sample between each bin of width $500$ pc according to the mean distance estimate.}
    \label{fig:ebv_distance_both}
\end{figure*}

 \begin{figure*}
    \centering
    \includegraphics[width=\textwidth]{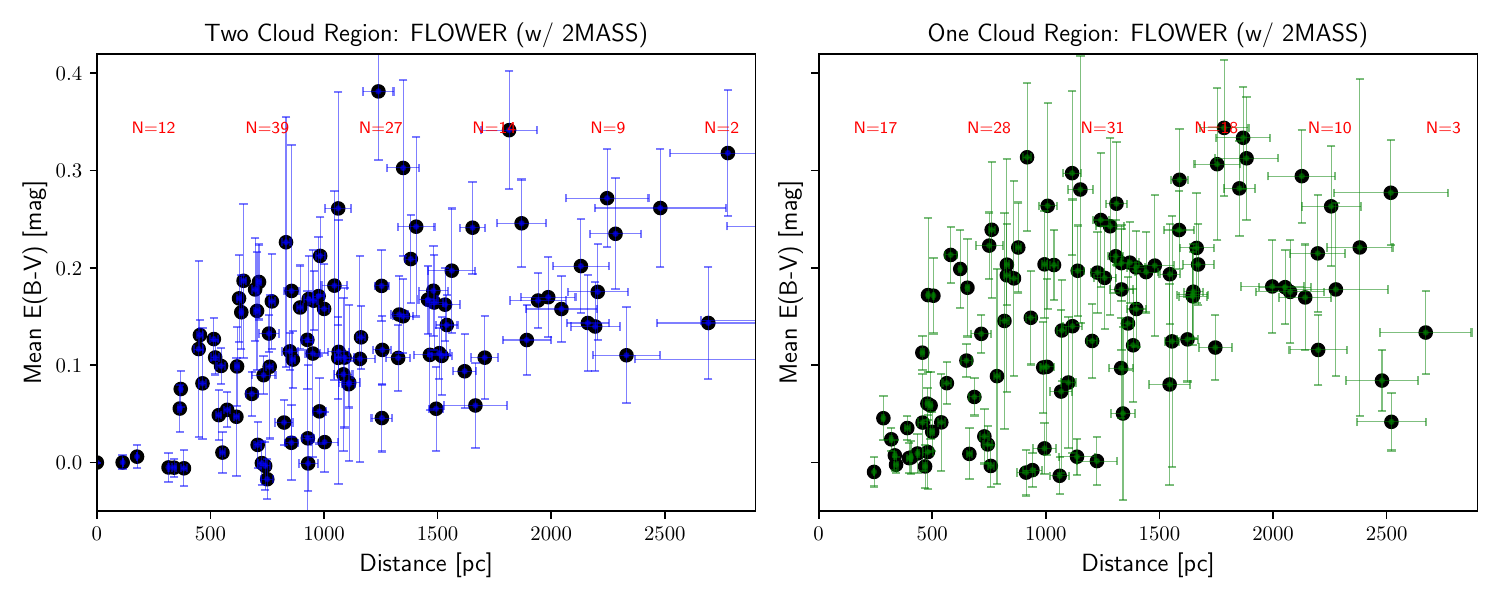}
        \caption{Gaia+PS1+2MASS: The FLOWER posterior mean reddening versus the posterior mean distance for the two-cloud (left) and one-cloud (right) regions. It is important to note that our zero point was adopted from the negligible extinction training sample and should not agree in overall magnitude with other methods exactly. In this plot, we show our method using more informative data but with less spatial resolution due to star count limits. The numbers in red $N=j$ indicate the number of stars in the sample between each bin of width $500$ pc according to the mean distance estimate.}
    \label{fig:ebv_distance_both_2mass}
\end{figure*}

 \begin{figure*}
    \centering
    \includegraphics[width=\textwidth]{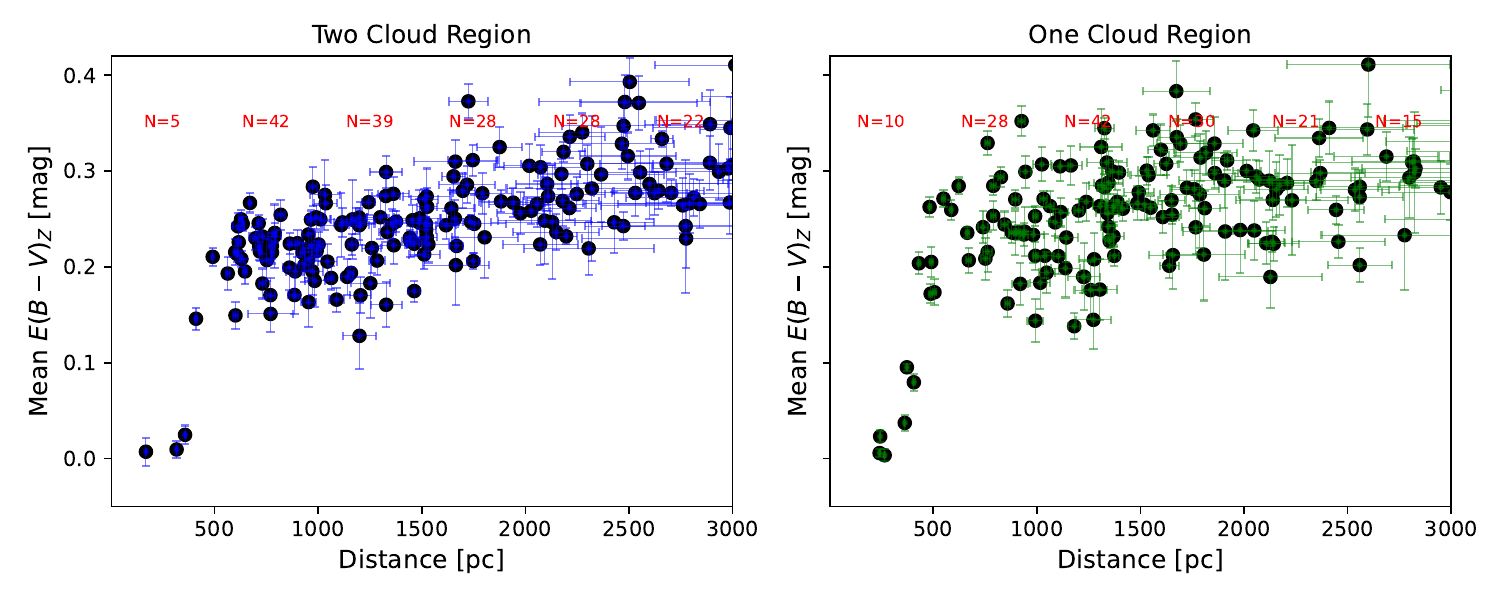}
        \caption{The \citealt{mlgreen} mean reddening, $E(B-V)_Z$ versus the mean distance for the two-cloud (left) and one-cloud (right) regions. {We find that their posterior mean reddening and distance pairs are in agreement with the distance to the first cloud predicted by polarisation signal. Furthermore, the trend is somewhat linearly increasing as a function of distance in both regions, with a steadier increase in the two-cloud region. It is difficult to ascertain the distance to the second cloud by eye, and it is noteworthy that the uncertainty on the distance estimate to the second cloud from the polarisation signal is high (distance between 1250 - 2140 pc).} The numbers in red $N=j$ indicate the number of stars in the sample between each bin of width $500$ pc according to the mean distance estimate.}
    \label{fig:ebv_distance_green_zhang}
\end{figure*}

 \begin{figure*}
    \centering
    \includegraphics[width=\textwidth]{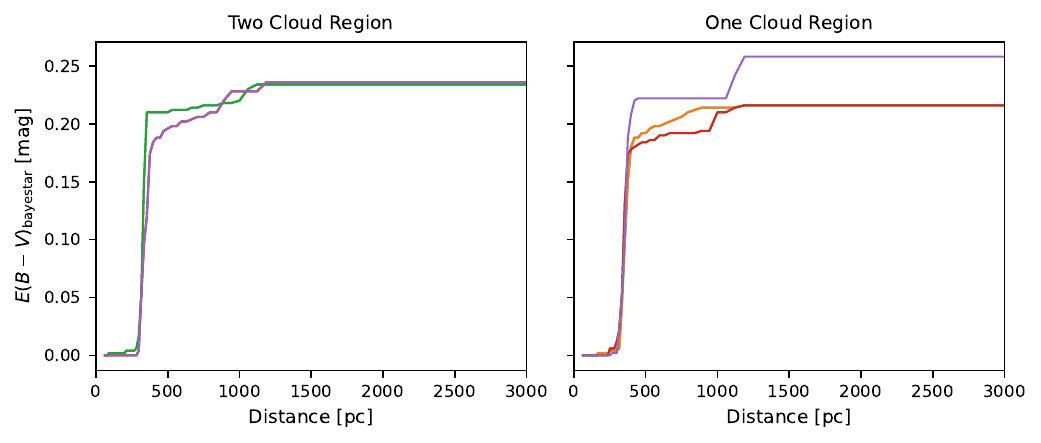}
        \caption{The \citealt{greenmap} mean reddening line-of-sight profile, $E(B-V)_{\rm bayestar}$ versus the distance bins for the two-cloud (left) and one-cloud (right) regions. Their posteriors assume spatial structure and are the values used in the \citealt{mlgreen} training. To generate these plots, we take 5 different lines-of-sight within the $0.16 ^\circ$ target region and overlay the posterior mean reddening from bayestar. Interestingly, we only find evidence of a second cloud in the one cloud region. The second cloud seen in polarisation may be too difficult to infer on the resolution of bayestar due to the low reddening nature of the second cloud.}
    \label{fig:bayestar_ebv}
\end{figure*}

\begin{figure*}
    \centering
    \includegraphics[width=\textwidth]{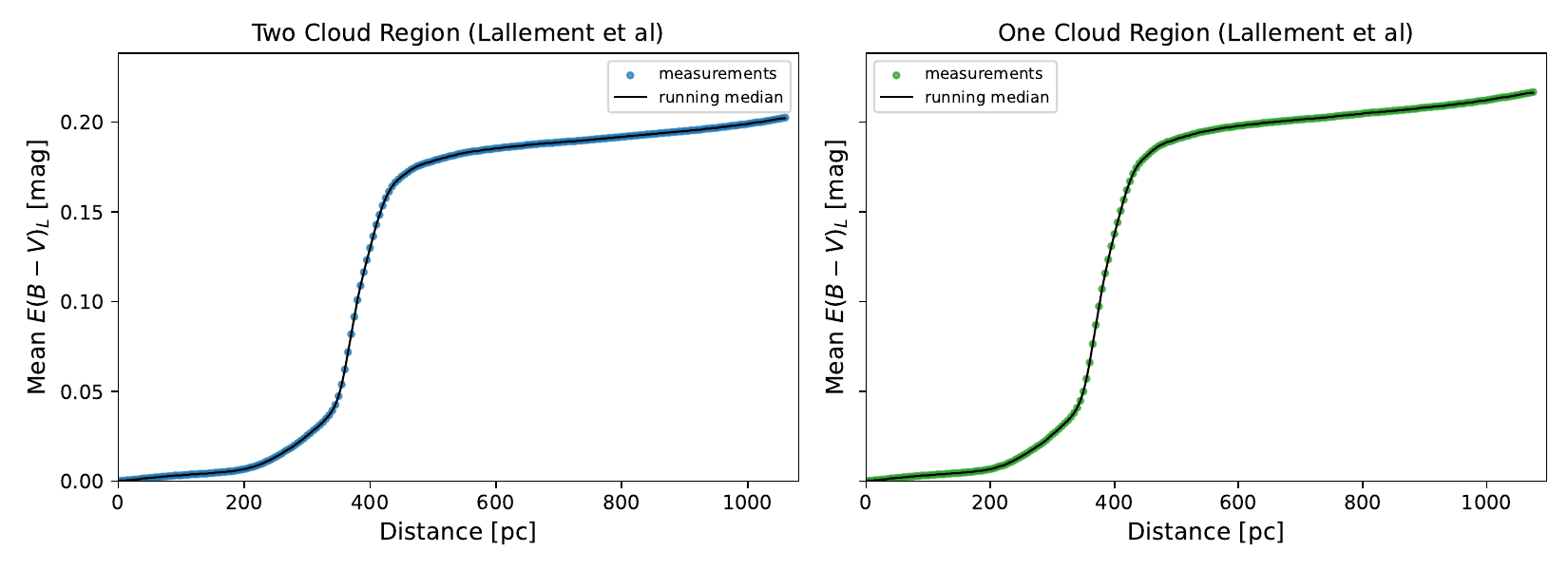}
        \caption{The \citealt{Lallement} mean reddening line-of-sight profile, $E(B-V)_{\rm L}$ versus the distance bins for the two-cloud (left) and one-cloud (right) regions. At these latitudes, their extinction only reaches 1.2 kpc.}
    \label{fig:lallement}
\end{figure*}

\section{Conclusions}
We introduce a method for accurately and precisely inferring photometric extinctions of stars at mid-to-high Galactic latitude $(|b|>20^\circ)$ by using probabilistic machine learning to model the distribution of photometry for zero-extinction stars in these regions. This method builds on the pioneering research in data-driven reddening calculations by learning the colour-magnitude relations as a probabilistic surface. This allows explicitly for the systematic change in stellar populations as a function of Galactic cylindrical coordinates and avoids the difficulties in mapping stellar parameters and their priors to the colour-magnitude diagram. Our method was developed to be independent of the photometric survey used, but requires an understanding of the selection function. This has had significant development in the context of the Gaia survey for the parent DR3 catalogue.

We trained a normalising flow on a sample of high Galactic latitude stars chosen from querying the \citealt{Lenz_2017} dust map to learn the conditional probability distribution of the photometric colour-magnitude vector of zero-extinction stars given Galactic cylindrical coordinates and absolute magnitude within $2.5$ kpc. Once we learnt the necessary distribution, we evaluated the effectiveness of our method at recovering reliable extinction posteriors using data from the Gaia, PAN-STARRS, and 2MASS surveys by adding synthetic extinction to real stars. We found that our method could reliably return the reddening posterior and provided state-of-the-art precision while retaining accuracy.

We evaluated our method for inferring reddening in two regions where the line-of-sight dust components had been resolved using polarisation measurements. We found that we could resolve both the large-scale structure and the low variation within our error bars. Moreover, we found that reddening estimates from other surveys failed to accurately constrain the fine-scale structure, even though their quoted errors on their reddening estimates should constrain the variation of the cloud. 

\subsection{Future work: Gaia BP/RP spectra and more photometric surveys}
In future, we will add Gaia BP/RP spectral data and more photometric surveys to provide a more accurate method for inferring the reddening. BP/RP spectra were not considered in this paper due to the technical challenge in quantifying their systematic errors.   Moreover, adopting a full extinction law in the forward process will introduce systematics that we need to be careful to account for in the modelling step. 
\subsection{Future work: variational inference methods}
In the problem at hand, where we are getting extinction estimates to measure small-area lines of sight, Hamiltonian Monte Carlo methods work very well at getting fast and reliable posterior inference. However, scaling this to all regions at high Galactic latitudes poses computational difficulties, particularly if we extend our data to use BP/RP spectra or other high-dimensional data. In the future, we will look to using variational inference to solve this problem at scale, but work needs to be carried out to minimise the error from the inference step and ensure that the variational inference posterior of reddening aligns sufficiently with the posteriors generated in this paper. Variational inference methods have been used before successfully for dust inference (see \citealt{2024MNRAS.535.2306U}, for an example).

\subsection{Future work: generating dust maps}
Future work will attempt to generate probabilistically robust dust maps that are consistent over small scales. Admittedly, this is a difficult task due to the assumptions needed currently to generate useful dust maps. However, we believe that the methods in this paper, which work towards a more robust single-star reddening posterior, are an important future step in this direction. 

\section*{Acknowledgements}
MOC is supported by the Gianna Angelopoulos Programme for Science, Technology and Innovation (GAPSTI) through the Science and Technology Facilities Council (STFC) studentship for astronomy.
GG acknowledges support from The Leverhulme Trust, through Emeritus Fellowship 2025-007.
KSM is supported by the European Union’s Horizon 2020 research and innovation programme under European Research Council Grant Agreement No 101002652 (BayeSN) and Marie Skłodowska-Curie Grant Agreement No 873089 (ASTROSTAT-II).

\section*{Data Availability}

This study uses data from Gaia DR3 \citep{GAIADR3}, 2MASS \citep{2MASS}, and PS1 \citep{PS1}, all of which are publicly available. Moreover, the stellar parameter and reddening estimates from \citealt{mlgreen} and the dust map from bayestar \citep{greenmap} are publicly available online. The code accompanying this paper can be found at https://github.com/mattocallaghan/FLOWER.



\bibliographystyle{mnras}
\bibliography{example} 




\appendix

\section{{Marginal Likelihood} Derivation} \label{derivation}
In this section, we derive the marginal likelihood loss function for a single star. We recall that the single-star contribution to the loss function is 
\begin{equation}\label{derivation_begin}
\begin{aligned}
p(\hat{\boldsymbol{\xi}}_s \mid S_s=1, \boldsymbol{\theta},\hat l_s,\hat b_s) 
&= \frac{1}{
P(S_s=1 \mid \boldsymbol{\theta} , \hat l_s, \hat b_s)
}\\&\quad\times
\displaystyle \int \! p(\hat{\boldsymbol{\xi}}_s \mid  \bm{x}_s, E_s=0,\mu_s) \,
\\ & \quad\times \,
p_{\boldsymbol{\theta}}(\bm{x}_s \mid \boldsymbol{\Phi}(d_s, \hat l_s, \hat b_s))
 \\
& \quad \times
I(M_s|\,\mu_s)p(d_s \mid \hat{l}_s, \hat{b}_s) \,\\&\quad \times
P(S_s=1 \mid  \boldsymbol{m_s}^{\rm unext}, \hat l_s,  \hat b_s)\\&\quad
 \, d\boldsymbol{x}_s \, dd_s  .
\end{aligned}
\end{equation}
and \begin{equation}
\begin{aligned}
P(S_s=1 \mid \boldsymbol{\theta} , \hat l_s,  \hat b_s) &= \int P(S_s=1 \mid \boldsymbol{m}^{\rm unext}_s, \hat l_s,\hat b_s) \\ & \quad\times 
p_{\boldsymbol{\theta}}(\bm{x}_s \mid \boldsymbol{\Phi}(d_s, \hat l_s, \hat b_s))
 \\
& \quad \times
I(M_s|\,\mu_s)p(d_s \mid \hat{l}_s, \hat{b}_s) \,\\&\quad
 \, d\bm{x}_s \, dd_s  .
\end{aligned}
\end{equation}
This integral is, in general, intractable. We instead use importance sampling to generate an estimate of the marginal likelihood for a single star. We introduce two importance sampling distributions $q(d|\hat \omega,\sigma_\omega,\hat l, \hat b)$ and $q(\boldsymbol{x}|\, \boldsymbol{\hat m}, \boldsymbol{\Sigma} , d)$ which simply sample around the data-estimated values of $\boldsymbol{x}$ and $\boldsymbol{d}$. {We choose $q(d|\hat \omega,\sigma_\omega,\hat l, \hat b)=N(d\mid \bar d, s_d^2)$ where $\bar d$ is the \citealt{Bailer_Jones_distance} posterior mean distance estimate and $s_d$ is the quoted central $68\%$ credible interval. Given a distance estimate, $q(\boldsymbol{x}|\, \boldsymbol{\hat m}, \boldsymbol{\Sigma} , d)=N({\boldsymbol{x}}|\,\hat{\boldsymbol{x}},\boldsymbol{\Sigma})$, where $\hat{\boldsymbol{x}}$ are the estimated unextinguished absolute magnitude and intrinsic colours calculated from the observed apparent magnitudes for the given distance $d$, assuming the photometry is free of dust extinction: $ \boldsymbol{\hat x}=\boldsymbol{\Lambda}(\boldsymbol{\hat m}-\boldsymbol{1}\mu(d))$.}

The benefit of using these distributions is that their density is almost identical to the likelihood components of the apparent magnitude and parallax, respectively. 
\begin{equation}
\begin{aligned}
p(\hat{\boldsymbol{\xi}}_s \mid S_s=1, \boldsymbol{\theta},\hat l_s,\hat b_s) 
&= \frac{1}{
P(S_s=1 \mid \boldsymbol{\theta} , \hat l_s, \hat b_s)
}\mathbb{E}_{q(d{_s}|\hat \omega{_s},{(\sigma_\omega)_s},\hat l{_s}, \hat b{_s})} \Big [\\&\quad
\displaystyle \mathbb{E}_{q(\boldsymbol{x}{_s}|\, \boldsymbol{\hat m}{_s}, \boldsymbol{\Sigma}{_s} , d{_s})}\Big [p(\hat{\boldsymbol{\xi}}_s \mid  \bm{x}_s, E_s=0,\mu_s)
\\ & \quad\times  \,
p_{\boldsymbol{\theta}}(\bm{x}_s \mid \boldsymbol{\Phi}(d_s, \hat l_s, \hat b_s))
 \\
& \quad \times
I(M_s|\,\mu_s)p(d_s \mid \hat{l}_s, \hat{b}_s) \,\\&\quad \times
P(S_s=1 \mid \boldsymbol{{ m_s}}{^{\rm unext}}, \hat l_s,  \hat b_s)\\
&\quad \times \frac{1}{q(d{_s}|\hat \omega{_s},{(\sigma_\omega)_s},\hat l{_s}, \hat b{_s})}\\
&\quad \times \frac{1}{q(\boldsymbol{x}{_s}|\, \boldsymbol{\hat m}{_s}, \boldsymbol{\Sigma}{_s} , d{_s})}
\Big]\Big]
\end{aligned}
\end{equation}
We chose the importance sampling distributions so that \begin{equation}
    p(\hat{\boldsymbol{\xi}}_s \mid  \bm{x}_s, E_s=0,\mu_s)/(q(d{_s}|\hat \omega{_s},{(\sigma_\omega)_s},\hat l{_s}, \hat b{_s}{)}q(\boldsymbol{x}{_s}|\, \boldsymbol{\hat m}{_s}, \boldsymbol{\Sigma}{_s} , d{_s}))\approx 1 \end{equation}
    
as evaluated densities. Therefore, the above loss function becomes 
\begin{equation}
\begin{aligned}
p(\hat{\boldsymbol{\xi}}_s \mid S_s=1, \boldsymbol{\theta},\hat l_s,\hat b_s) 
&\approx \frac{1}{
P(S_s=1 \mid \boldsymbol{\theta} , \hat l_s, \hat b_s)
}\mathbb{E}_{q(d{_s}|\hat \omega{_s},{(\sigma_\omega)_s},\hat l{_s}, \hat b{_s})} \Big [\\&\quad
\displaystyle \mathbb{E}_{q(\boldsymbol{x}{_s}|\, \boldsymbol{\hat m}{_s}, \boldsymbol{\Sigma}{_s} , d{_s})}\Big [p_{\boldsymbol{\theta}}(\bm{x}_s \mid \boldsymbol{\Phi}(d_s, \hat l_s, \hat b_s))
 \\
& \quad \times
I(M_s|\,\mu_s)p(d_s \mid \hat{l}_s, \hat{b}_s) \,\\&\quad \times
P(S_s=1 \mid  \boldsymbol{m{_s}}{^{\rm unext}}, \hat l_s,  \hat b_s)
\Big]\Big]
\end{aligned}
\end{equation}
Finally, we must evaluate the term $P(S_s=1 \mid \boldsymbol{\theta}, \hat l_s, \hat b_s)$. It should be noted that {a} significant analysis of the selection function should be carried out {before} training so as not to waste significant time implementing a more complicated loss function than necessary. With sensible cuts, one can construct a selection function that is nearly independent of the sky position{,} and the selection function normali{z}ation is almost uniform for every star. In this situation{,} the following term may become redundant. It is not necessary to calculate this integral with the same importance sampling distributions. Moreover, if the selection function is expected to be relatively uniform over the parameter space generated by the selection function{,} then adopting the importance sampling distributions can provide a high variance estimate of this term{,} so extra care should be taken. However, we opt to use the same importance sampling distributions{,} and the term becomes 
\begin{equation}
\begin{aligned}
P(S_s=1 \mid \boldsymbol{\theta} , \hat l_s,  \hat b_s) &= \mathbb{E}_{q(d{_s}|\hat \omega{_s},{(\sigma_\omega)_s},\hat l{_s}, \hat b{_s})} \Big [\\&\quad
\displaystyle \mathbb{E}_{q(\boldsymbol{x}{_s}|\, \boldsymbol{\hat m}{_s}, \boldsymbol{\Sigma}{_s} , d{_s})}\Big [P(S_s=1 \mid \boldsymbol{m}{^{\rm unext}}_s, \hat l_s,\hat b_s)
\\ & \quad\times  \,
p_{\boldsymbol{\theta}}(\bm{x}_s \mid \boldsymbol{\Phi}(d_s, \hat l_s, \hat b_s))
 \\
& \quad \times
I(M_s|\,\mu_s)p(d_s \mid \hat{l}_s, \hat{b}_s) \\
&\quad \times \frac{1}{q(d{_s}|\hat \omega{_s},{(\sigma_\omega)_s},\hat l{_s}, \hat b{_s})}\\
&\quad \times \frac{1}{q(\boldsymbol{x}{_s}|\, \boldsymbol{\hat m}{_s}, \boldsymbol{\Sigma}{_s} , d{_s})}
\Big]\Big]
\end{aligned}
\end{equation}
Then, from samples of the importance distributions, we can take the natural logarithm of Equation \ref{derivation_begin} and obtain a noisy estimate of the single-star marginal likelihood. The final loss function is obtained by summing {over} each of these terms.
\bsp	
\label{lastpage}
\end{document}